\documentclass[useAMS,usenatbib]{mn2e}
\usepackage{times,graphicx}
\usepackage[caption=false]{subfig}
\usepackage{morefloats}
\usepackage{natbib}
\usepackage{longtable,lscape,supertabular}
\usepackage{booktabs}
\usepackage{url}
\usepackage{paralist}

\input{epsf}

\title[Mid-IR Spectroscopy of PNe]
{
\textit{Spitzer} mid-infrared spectroscopic observations of planetary 
nebulae\thanks{
This work is based on observations made with the \textit{Spitzer} Space 
Telescope, which is operated by the Jet Propulsion Laboratory at the 
California Institute of Technology under a contract with NASA.
} 
}

\author[H. Mata et al.] 
{H.\ Mata$^{1}$\thanks{E-mail:
gerardo@astro.iam.udg.mx (GRL)},
G.\ Ramos-Larios$^{2}$, 
M.A.\ Guerrero$^{3}$, 
A.\ Nigoche-Netro$^{2}$, 
J.A.\ Toal\'a$^{3,4}$, 
\newauthor
X.\ Fang$^{3}$,
G.\ Rubio$^{1}$,  
S.N.\ Kemp$^{2}$, 
S.G.\ Navarro$^{2}$, and 
L.J.\ Corral$^{2}$ \\
$^{1}$Centro Universitario de Ciencias Exactas e Ingenier\'{\i}as, 
Universidad de Guadalajara, Av. Revoluci\'on 1500, Guadalajara, 
Jalisco, Mexico\\
$^{2}$Instituto de Astronom\'{\i}a y Meteorolog\'{\i}a, CUCEI, 
Universidad de Guadalajara, Av.\ Vallarta No.\ 2602, Col.\ Arcos 
Vallarta, 44130 Guadalajara, Jalisco, Mexico\\
$^{3}$Instituto de Astrof\'\i sica de Andaluc\'\i a, IAA-CSIC, 
C/ Glorieta de la Astronom\'{\i}a s/n, 18008 Granada, Spain\\
$^{4}$Institute of Astronomy and Astrophysics, Academia Sinica (ASIAA),
Taipei 10617, Taiwan
}

\begin{document}

\date{Received 2015 May 09; in original form 2014 August 18}

\pagerange{\pageref{firstpage}--\pageref{lastpage}} \pubyear{2016}

\maketitle

\label{firstpage}

\begin{abstract}

We present {\it Spitzer Space Telescope} archival mid-infrared (mid-IR) spectroscopy of a sample of eleven planetary nebulae (PNe). The observations, acquired with the \textit{Spitzer} Infrared Spectrograph (IRS), cover the spectral range 5.2-14.5 $\mu$m	that includes the H$_{2}$ 0-0 S(2) to S(7) rotational emission lines. This wavelength coverage has allowed us to derive the Boltzmann distribution and calculate the H$_2$ rotational excitation temperature (T$_\mathrm{ex}$). 
The derived excitation temperatures have consistent values $\simeq$900$\pm$70 K for different sources despite their different structural components.  
We also report the detection of mid-IR ionic lines of [Ar~{\sc iii}], 
[S~{\sc iv}], and [Ne~{\sc ii}] in most objects, and polycyclic aromatic 
hydrocarbon (PAH) features in a few cases.  
The decline of the [Ar~{\sc iii}]/[Ne~{\sc ii}] line ratio with the 
stellar effective temperature can be explained either by a true neon 
enrichment or by high density circumstellar regions of PNe that 
presumably descend from higher mass progenitor stars.

\end{abstract}


\section{Introduction}

\label{Intro}

Molecular hydrogen (H$_2$) is the most abundant molecule in the 
Universe.  
Its excitation and emission has been thoroughly studied theoretically and 
the emission line ratios have been modelled in detail (\citealp{Burton1992b}; 
\citealp{Burton1992a}).

Transitions from the rotational and vibrational levels of the ground 
electronic state of the H$_2$ molecule produce emission lines in the 
spectral range from the ultraviolet (UV) to the far-infrared (far-IR).  
The most relevant excitation mechanisms of H$_2$ are:
\begin{inparaenum}[\itshape i\upshape)]
\item 
fluorescence or radiative pumping through the absorption of UV radiation 
from a hot source, and 
\item 
collisional excitation. 
\end{inparaenum}  
Photodissociation regions (PDRs), where a hot source of UV photons irradiates 
neutral clouds of material, are typical sites for UV excitation of H$_2$.  
This includes star-forming regions, but it also applies to any cloud 
illuminated by a local interstellar radiation field (\citealp{Burton1992b}; 
\citealp{Habart}), and particularly the photodissociation front around a 
planetary nebula (PN) central star (\citealp{Black}; \citealp{Dinerstein}; 
\citealp{Sternberg}).  
On the other hand, shocks propagating into a medium can heat the 
gas and collisionally excite the H$_2$ molecules 
\citep[e.g.,][]{HollenbachShull1977, Draine1983, Smith1995}.  
This is also the case for the outflowing gas in PNe, which can produce 
the collisional excitation of H$_2$ in shock regions even if they are 
embedded in the ionized regions of PNe with high temperature central 
stars \citep{AlemanGruenwald}.

Planetary nebulae (PNe) are classical targets for the study of molecular 
hydrogen, very particularly through its infrared line emission.  
Near-infrared (near-IR) narrowband imaging surveys of PNe have detected the 
presence of H$_2$ in many objects 
\citep[e.g.,][]{Storey1984, Webster1988, Latter1995, Kastner, Bohigas, Bohigasa, Guerrero2000}.  
The H$_2$ emission is detected in the envelopes and PDRs of PNe, where 
the molecule can be shielded from the UV radiation of the central star 
by the inner gas and dust or by high-density clumps \citep{Aleman}.  
Using the known general behaviour and properties of H$_2$ in the 
interstellar medium (ISM; \citealp{Habart}), several studies have 
attempted to gain insight into the conditions in PNe where this 
species is found.  
The correlation between the bipolar morphology of PNe and the H$_2$ 
detection is well documented \citep{Kastner}, suggesting that the 
H$_2$-bright waists of bipolar PNe are the remnants of dense, 
molecule-rich circumstellar disks.  
Alternatively, it has been claimed that the preferential detection of 
H$_2$ emission in bipolar PNe may be an effect of the typically higher 
effective temperature of their central stars (\citealp{Phillips2006}; 
\citealp{Aleman}; \citealp{AlemanGruenwald}). 
The consensus is that the brightest H$_2$ emitters tend to be bipolar 
PNe, but deep near-IR imaging of the 2.122 $\mu$m H$_2$ emission line 
shows that this emission is not exclusive to this morphology 
\citep{Marquez}.

Besides the excellent near-IR spectroscopic survey of PNe and study of 
upper vibrational levels carried out by \citet{Hora1999}, the actual 
investigation of the excitation mechanism of H$_2$ in PNe and the 
calculations of its excitation temperature have been made in the context 
of other detailed studies \citep[see e.g.,][and references therein]{Hora1994, 
Hora1996, Shupe1998, Rudy, Lumsden, GarciaHernandez2002, Davis, Likkel2006, 
Ramos, Aleman2011}. 
Mid-infrared (mid-IR) spectroscopy is a powerful tool for the study of the 
thermal and molecular emission from PNe, since such observations are much 
less affected by dust extinction. 
Mid-IR studies of PNe have focused on imaging surveys \citep[e.g.,][]{
Hora2004, Phillips2008, Zhang2009, Phillips2010, Zhang2012}, but mid-IR 
spectroscopy of PNe has only been carried out for individual objects 
for which chemical abundances and H$_2$ excitation temperatures were 
obtained \citep[][]{ Cox, Matsuura2005, Hora2006, Phillips2011}.  
Currently, mid-IR studies of the excitation temperature of H$_2$ 
in PNe are scarce.

\begin{table*}
\centering
\begin{minipage}{160 mm}

\caption{Basic properties of target PNe.}\label{GeneralObjectData}
\begin{tabular}{@{}lccccccccl@{}}
\toprule
Object	& 
\multicolumn{1}{c}{R.A.} & 
\multicolumn{1}{c}{Dec.} & 
Distance	& Angular Size	& Physical Size	& log N$_\mathrm{e}$ & V$_\mathrm{exp}$ & T$_\mathrm{eff}$ & References \\ 
       & 
\multicolumn{2}{c}{(J2000)} & 
\multicolumn{1}{c}{(kpc)}    & 
\multicolumn{1}{c}{($^{\prime\prime}$)} & 
\multicolumn{1}{c}{(pc$^{2}$)}       & 
\multicolumn{1}{c}{(cm$^{-3}$)}	& 
\multicolumn{1}{c}{(km s$^{-1}$)} & (10$^3$~K)  & \\ \midrule
M\,2-51   & 22 16 03.3 & $+$57 28 41 & 1.60 & 83$\times$58    & 0.64$\times$0.45 & 2.94 & $\cdots$  & $\cdots$ & 1, 2 \\
NGC\,2346 & 07 09 22.1 & $-$00 48 17 & 0.90 & 120$\times$65   & 0.52$\times$0.28 & 2.64 & 12        & \textgreater 80 & 3, 4 \\
NGC\,2440 & 07 41 55.4 & $-$18 12 33 & 1.90 & 74$\times$45    & 0.68$\times$0.42 & 3.06 & 25        & 208 & 4      \\
NGC\,2818 & 09 16 00.5 & $-$36 37 32 & 2.84 & 110$\times$55   & 1.52$\times$0.75 & 3.11 & 18.5      & 160 & 5, 6, 7, 8 \\
NGC\,3132 & 10 07 01.8 & $-$40 26 10 & 0.81 & 85$\times$58    & 0.33$\times$0.23 & 2.80 & 21        & 100 & 4         \\
NGC\,6072 & 16 12 58.6 & $-$36 13 49 & 1.39 & 120$\times$70   & 0.81$\times$0.47 & 2.64 & 10        & 140 & 4         \\
NGC\,6445 & 17 49 15.3 & $-$20 00 35 & 1.39 & 50$\times$30    & 0.33$\times$0.20 & 3.16 & 38        & 170 & 4        \\
NGC\,6537 & 18 05 13.4 & $-$19 50 13 & 2.00 & 230$\times$56   & 2.23$\times$0.54 & 4.11 & 18        & 250 & 4        \\
NGC\,6720 & 18 73 35.1 & $+$33 01 45 & 0.70 & 96$\times$74    & 0.33$\times$0.25 & 2.90 & 22        & 125 & 4   \\
NGC\,6781 & 19 18 28.1 & $+$06 32 19 & 0.95 & 200$\times$120  & 0.92$\times$0.55 & 2.41 & 12        & 112 & 4    \\
NGC\,7293 & 22 29 38.8 & $-$20 50 15 & 0.22 & 1150$\times$770 & 1.22$\times$0.82 & 1.86 & 21        & 110 & 4    \\ \bottomrule
\end{tabular}
\footnotesize{
Notes: \\
log N$_\mathrm{e}$: logarithm of electronic density; 
V$_\mathrm{exp}$: expansion velocity; 
T$_\mathrm{eff}$: central star effective temperature. 
The sizes shown in the table for each object refer to the main nebula only 
and do not consider the halo. \\
References:
(1) \citealp{Acker}; (2) \citealp{Jewitt}; (3) \citealp{Vicini}; (4) \citealp{Frew}; 
(5) \citealp{Ortiz}; (6) \citealp{Bohigasa}; (7) \citealp{Bohigasb}; (8) \citealp{Vazquez}.  
}
\end{minipage}
\end{table*}

Using archival \textit{Spitzer} mid-IR spectroscopy, we have conducted 
a spectroscopic survey of PNe to assess the presence of ionic species, 
polycyclic aromatic hydrocarbons (PAHs), and molecular hydrogen.  
The H$_2$ excitation temperature has been derived using its mid-IR 
emission lines.  
These temperatures shed light on the purely rotational component of 
the excitation temperatures to complement the rotational-vibrational 
studies obtained by near-IR observations.

Section \ref{ObsData} details the criteria for target selection and 
the data processing.  
Section \ref{Results} describes the mid-IR spectroscopic properties 
of each PN in our sample.  
A review of theory behind data analysis is given in section 
\ref{H2Excitation} and the final discussion is presented in 
section \ref{Discussion}.

\section{Archival Data: Selection and Reduction}
\label{ObsData}

\subsection{The {\it Spitzer} archive}

This study made use of the \textit{Spitzer Space Telescope} 
(\textit{Spitzer}; \citealp{Werner}) database. 
We used spectroscopic observations obtained with the Infrared 
Spectrograph (IRS; \citealp{Houck}), one of three \textit{Spitzer} focal plane instruments. It consists of four separate modules, namely Short-Low (SL), Short-High (SH), Long-Low (LL), and Long-High (LH), which provide low (R$\sim$60--128) and moderate (R$\sim$600) resolution spectroscopic capabilities from 5.2 $\mu$m to 38 $\mu$m. The low-resolution modules employ a long-slit that allows the acquisition of spatially-resolved spectral information on the same detector array.

In this study we will use low-resolution IRS spectra obtained with the Short-Low SL1 and SL2 modules. The aperture sizes for the SL1 and SL2 modules are 3.7$\times$57 arcsec$^2$ and 3.6$\times$57 arcsec$^2$, respectively. They cover different spectral ranges, 7.4-14.5 $\mu$m for the SL1 and 5.2-7.7 $\mu$m for the SL2. This spectral range is specially suited for analyses of Boltzmann distributions (see section \ref{H2Excitation}) as it includes the S(2) to S(7) rotational lines of H$_2$.

\subsection{Target Selection}

The procedure for the target selection in this study is as follows: \begin{inparaenum}[\itshape 1\upshape)]
\item we first searched through the literature for PNe that are known to exhibit H$_{2}$ emission; \item then we queried the {\it Spitzer} online database for objects with SL1 and SL2 spectra modules in the 5.2-14.5 $\mu$m range; and \item objects without suitable observations through nearby slits that 
could be used for background subtraction were discarded.
\end{inparaenum}

\begin{table*}
\begin{minipage}{100mm}
\centering
\caption{\textit{Spitzer} AORs used for the survey.}\label{AORs}
\begin{tabular}{@{}lrlrll@{}}
\toprule
Object   & 
\multicolumn{1}{c}{AORs}    & 
\multicolumn{1}{c}{Type}    & 
\multicolumn{1}{c}{Program} & 
\multicolumn{1}{c}{PI}      & 
\multicolumn{1}{c}{Date}    \\ \midrule
M\,2-51   & 4114688  & IRS Stare & 45      & Roellig, Thomas & 2004-01-06 \\
         & 4114944  &           &         &                 &            \\
NGC\,2346 & 17462272 & IRS Map   & 30482   & Houck, James R. & 2007-05-03 \\
         & 17462016 &           &         &                 &            \\
         & 17461248 &           &         &                 &            \\
NGC\,2440 & 27358208 & IRS Map   & 50179   & Sellgren, Kris  & 2008-12-05 \\
         & 27358464 &           &         &                 & 2004-12-04 \\
         & 27360512 &           &         &                 &            \\
NGC\,2818 & 27361280 & IRS Map   & 50179   & Sellgren, Kris  & 2009-01-06 \\
NGC\,3132 & 27358720 & IRS Map   & 50179   & Sellgren, Kris  & 2009-01-14 \\
NGC\,6072 & 4115200  & IRS Stare & 45      & Roellig, Thomas & 2005-08-11 \\
         & 4115456  &           &         &                 &            \\
NGC\,6445 & 27361024 & IRS Map   & 50179   & Sellgren, Kris  & 2008-10-18 \\
NGC\,6537 & 27358976 & IRS Map   & 50179   & Sellgren, Kris  & 2008-11-04 \\
NGC\,6720 & 15878656 & IRS Map   & 1424    & Calibration     & 2005-09-13 \\
NGC\,6781 & 16099072 & IRS Map   & 1425    & Calibration     & 2005-10-19 \\
NGC\,7293 & 13736192 & IRS Stare & 1421    & Calibration     & 2005-05-29 \\ \bottomrule
\end{tabular}
\end{minipage}

\end{table*}

Eventually, eleven objects were selected for analysis:  
M\,2-51, NGC\,2346, NGC\,2440, NGC\,2818, NGC\,3132, NGC\,6072, NGC\,6445, NGC\,6537, NGC\,6720, NGC\,6781, and NGC\,7293.  
Basic properties of our targets are presented in 
Table~\ref{GeneralObjectData}. The information on the available Astronomical Observation Requests (AORs) for each object is shown in Table~\ref{AORs}.  Several PNe known to show H$_2$ emission with available \textit{Spitzer} data did not meet the criteria above and were thus not included in this study.  
No suitable background information was available for NGC\,40, NGC\,650-1, NGC\,6369, and NGC\,7026, no H$_{2}$ emission was detected in the SL spectra of J\,900, NGC\,6629, and NGC\,7027, and no SL spectra were available for Hb\,12, Hu\,2-1, IC\,5117, and M\,2-9.  

\begin{figure*}
\vspace{5mm}
\centering
\begin{minipage}{.4\textwidth}
  \centering
  \includegraphics[width=.95\linewidth]{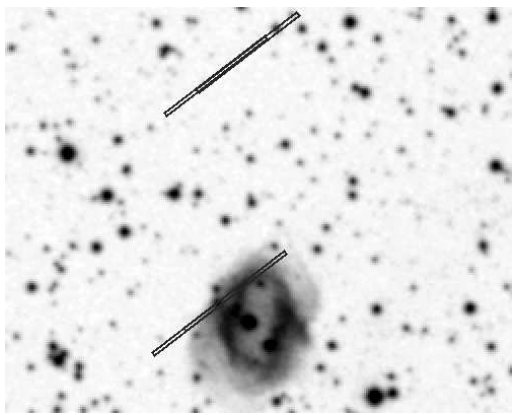}
  \label{fig:M2-51Slit}
\end{minipage}
\begin{minipage}{.5\textwidth}
  \centering
  \includegraphics[width=0.95\linewidth]{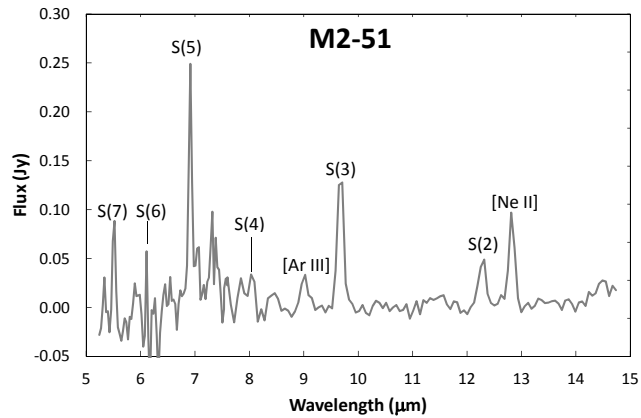}
  \label{fig:M2-51Flux}
\end{minipage}
\caption{
{\it (left)} DSS red image of M\,2-51 superimposed with the slit positions of the selected source and background (outside the nebula) SL spectra.  North is up and east to the left.  
The DSS red image of PNe is expected to be dominated by the H$\alpha$ and [N~{\sc ii}] nebular emission. {\it (right)} \textit{Spitzer} mid-IR (SL1 + SL2) spectrum of M\,2-51, where several H$_{2}$ and ionic lines are identified. 
}
\label{M2-51Spectra}
\end{figure*}

\subsection{Data Reduction}
\label{DataReduction}

After selecting the SL1 and SL2 bands for each nebula in 
Table~\ref{AORs}, the Basic Calibrated Data (BCD) were downloaded for analysis from the NASA/IPAC Archive.  
The \textit{Spitzer} BCD data have been dark-subtracted, flat field corrected, and wavelength and flux calibrated. The data were further processed using the software package 
{\sc CUBISM}\footnote{
{\sc CUBISM}, and a manual detailing its use, are available from the \textit{Spitzer} Science Center at http://ssc.spitzer.caltech.edu/archanaly/contributed/cubism}
(Cube Builder for IRS Spectral Mapping; \citealp{2007PASP..119.1133S}) to subtract the sky emission and produce clean, low-resolution spectra for each object. The various processing steps within {\sc CUBISM} include  characterization of noise in the data and removal of bad pixels. One-dimensional spectra were then extracted and line fluxes measured accounting for bad pixel correction and map creation (e.g. a continuum-subtracted line image).

\begin{figure*}
\vspace{5mm}
\centering
\begin{minipage}{.4\textwidth}
  \centering
  \includegraphics[width=.95\linewidth]{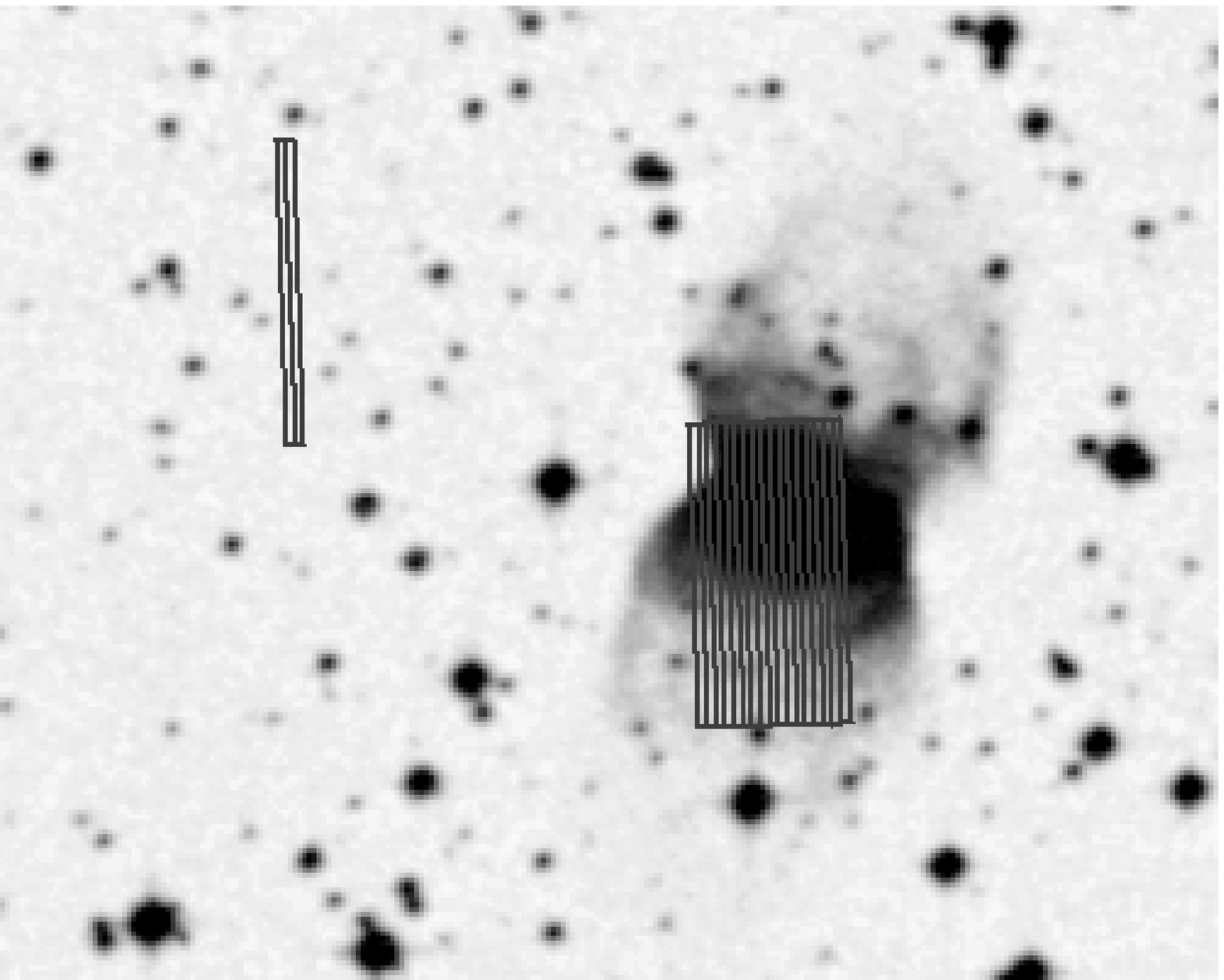}
  \label{fig:NGC2346Slit}
\end{minipage}
\begin{minipage}{.5\textwidth}
  \centering
  \includegraphics[width=0.95\linewidth]{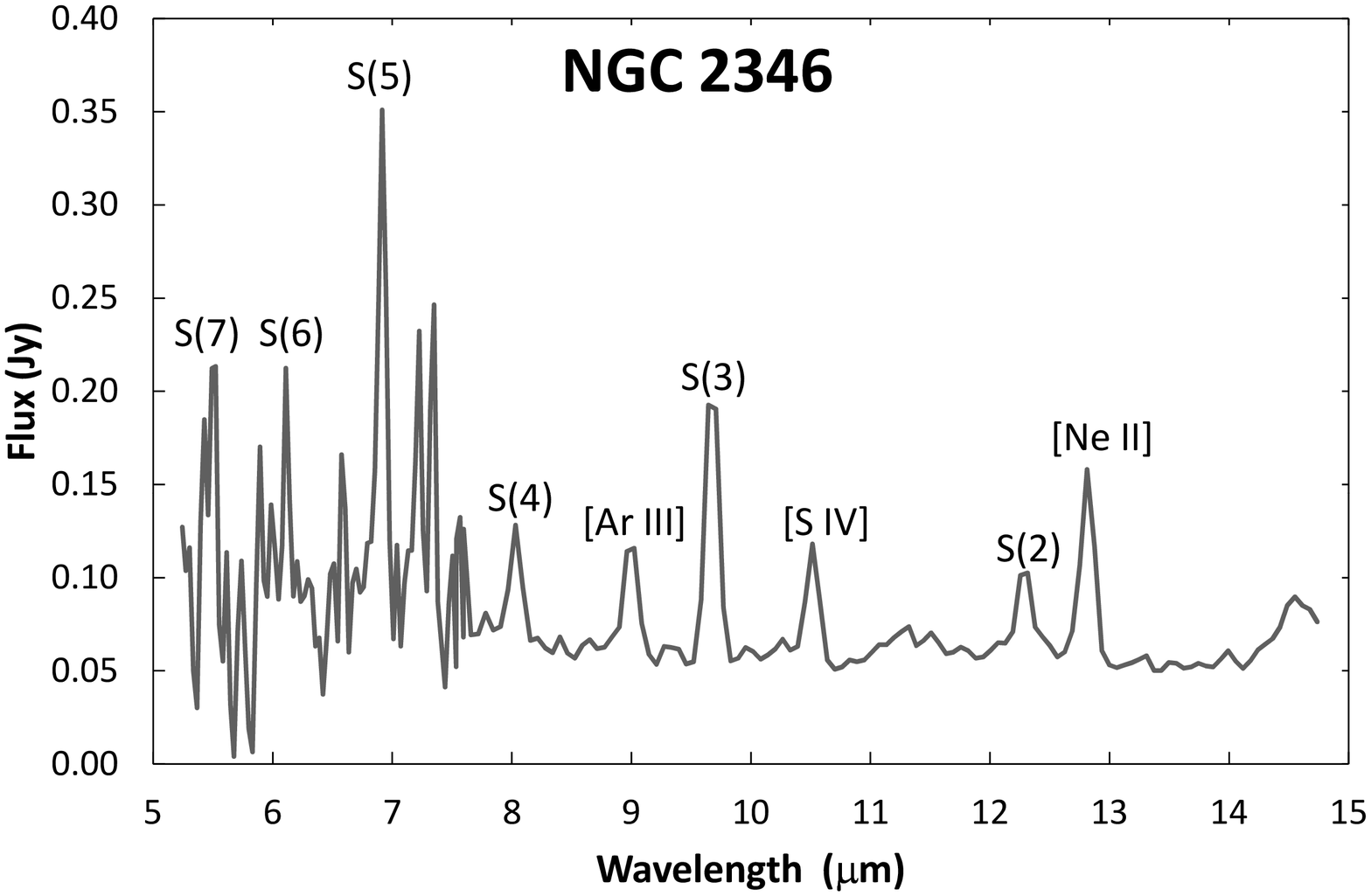}
  \label{fig:NGC2346Flux}
\end{minipage}
\caption{Same as Fig.1 but for NGC\,2346}
\label{NGC2346Spectra}
\end{figure*}

\begin{figure*}
\vspace{5mm}
\centering
\begin{minipage}{.4\textwidth}
  \centering
  \includegraphics[width=.95\linewidth]{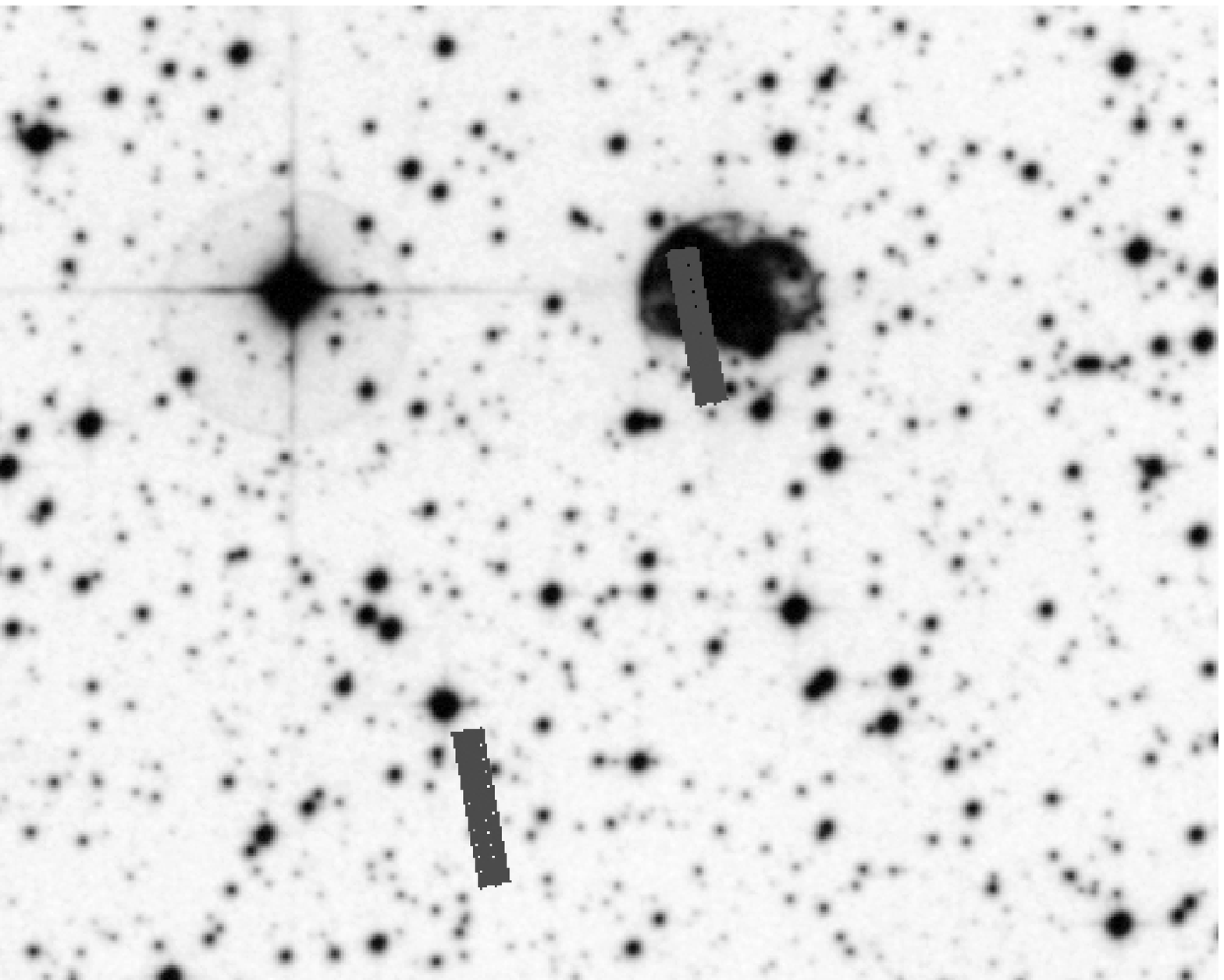}
  \label{fig:NGC2440Slit}
\end{minipage}
\begin{minipage}{.5\textwidth}
  \centering
  \includegraphics[width=0.94\linewidth]{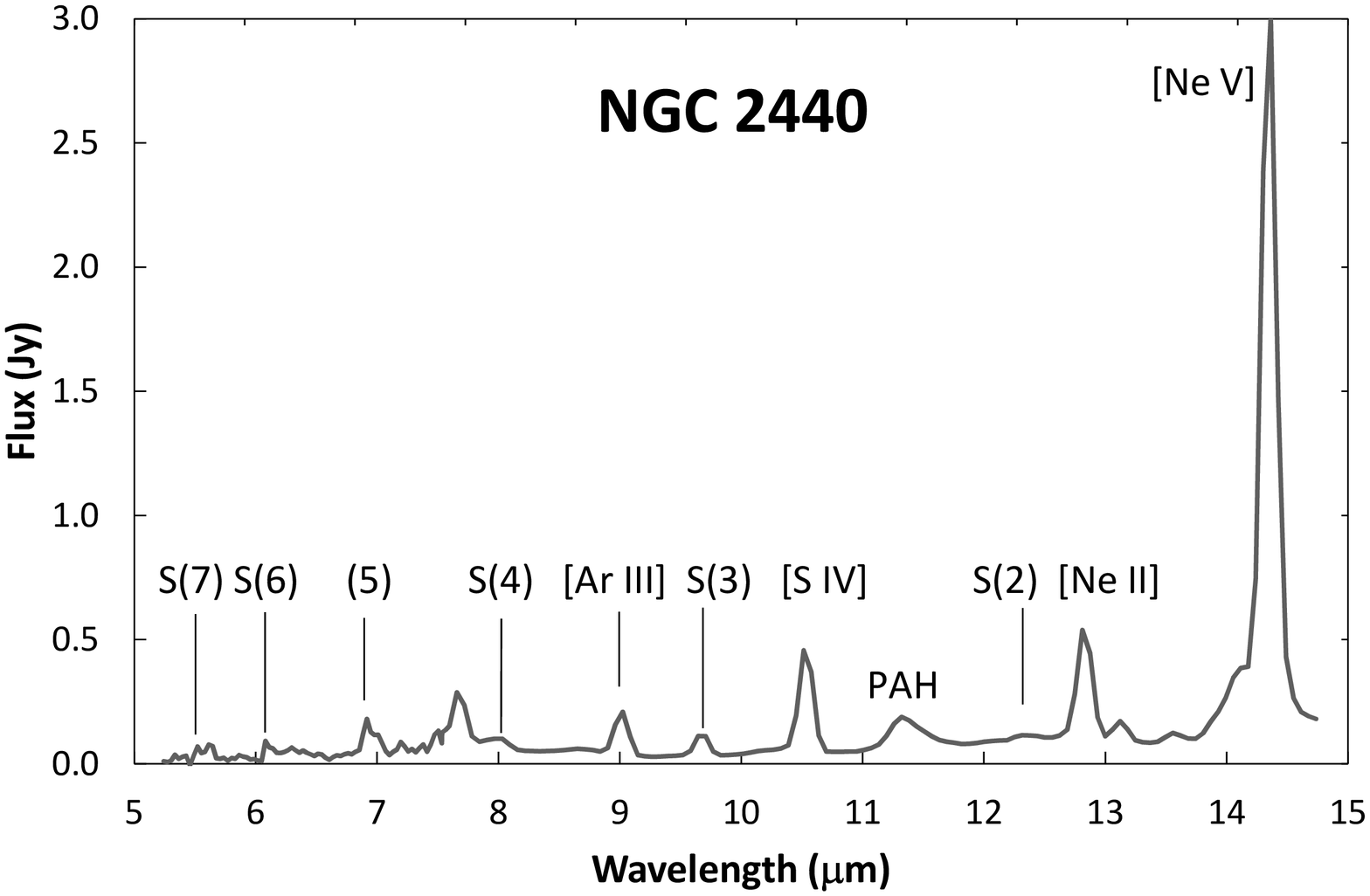}
  \label{fig:NGC2440Flux}
\end{minipage}
\caption{Same as Fig.1 but for NGC\,2440.}
\label{NGC2440Spectra}
\end{figure*}

\section{Results}
\label{Results}

\subsection{Slits and Spectra}

\begin{figure*}
\vspace{5mm}
\centering
\begin{minipage}{.4\textwidth}
  \centering
  \includegraphics[width=.95\linewidth]{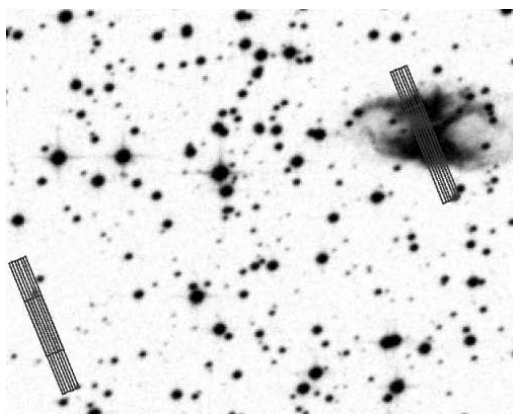}
  \label{fig:NGC2818Slit}
\end{minipage}
\begin{minipage}{.5\textwidth}
  \centering
  \includegraphics[width=0.96\linewidth]{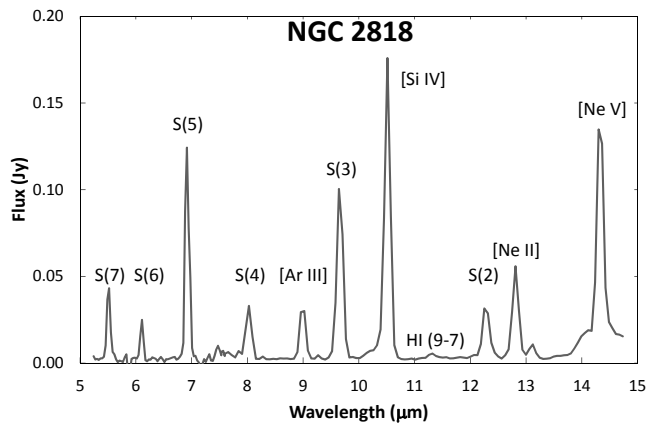}
  \label{fig:NGC2818Flux}
\end{minipage}
\caption{Same as Fig.1 but for NGC\,2818.}
\label{NGC2818Spectra}
\end{figure*}

\begin{figure*}
\vspace{5mm}
\centering
\begin{minipage}{.4\textwidth}
  \centering
  \includegraphics[width=.95\linewidth]{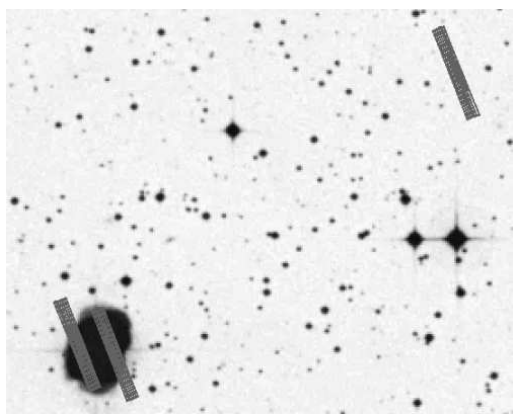}
  \label{fig:NGC3132Slit}
\end{minipage}
\begin{minipage}{.5\textwidth}
  \centering
  \includegraphics[width=0.94\linewidth]{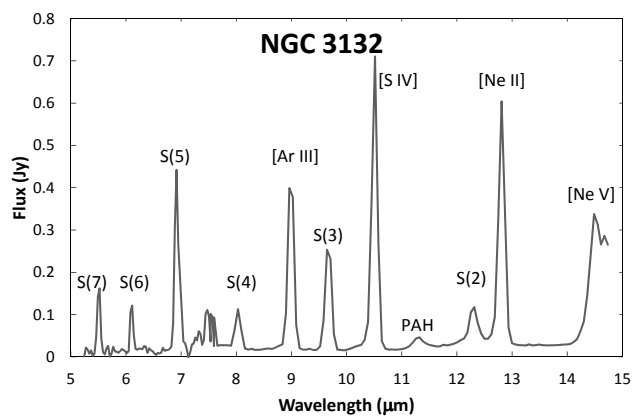}
  \label{fig:NGC3132Flux}
\end{minipage}
\caption{Same as Fig.1 but for NGC\,3132.}
\label{NGC3132Spectra}
\end{figure*}

\begin{figure*}
\vspace{5mm}
\centering
\begin{minipage}{.4\textwidth}
  \centering
  \includegraphics[width=.95\linewidth]{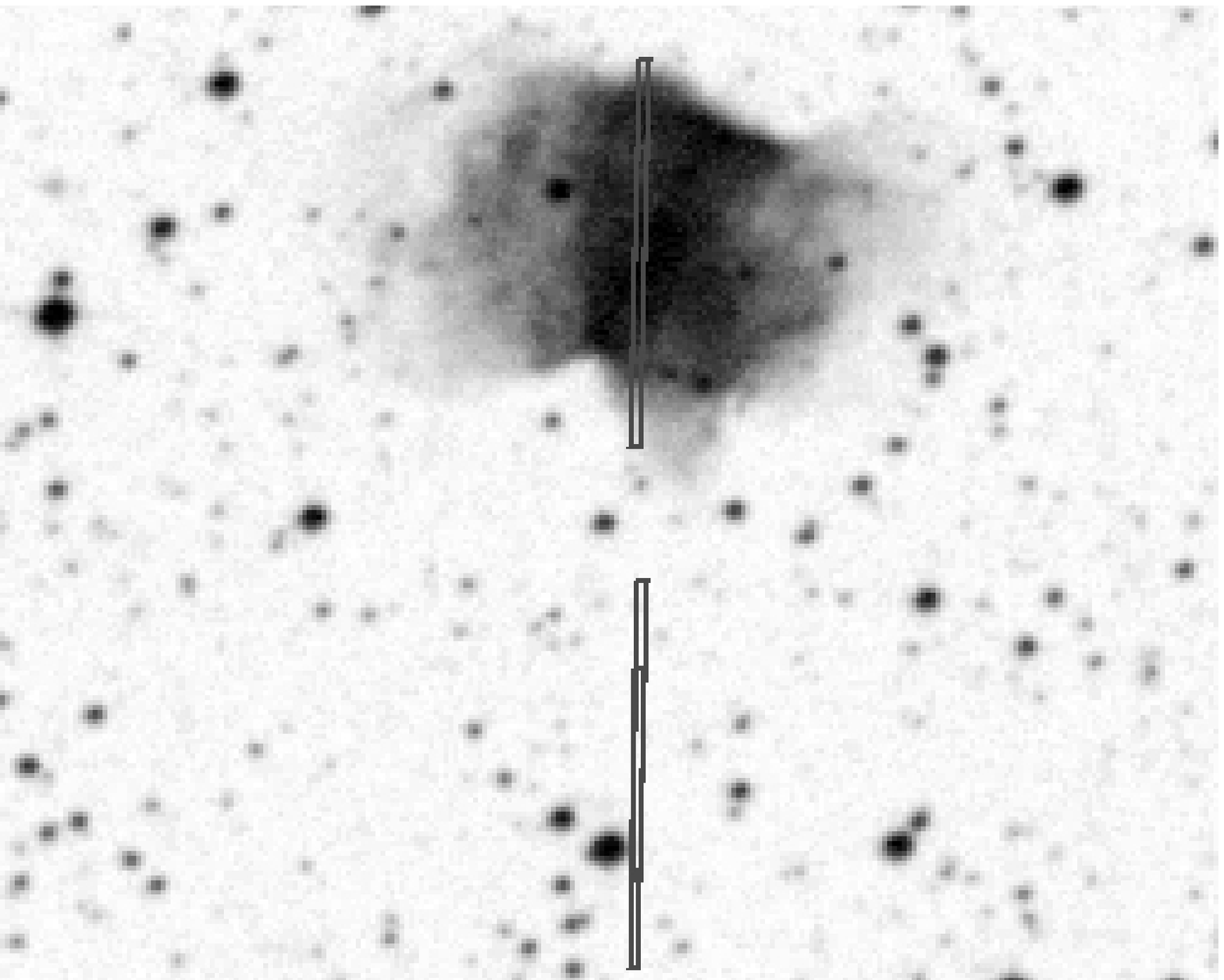}
  \label{fig:NGC6072Slit}
\end{minipage}
\begin{minipage}{.5\textwidth}
  \centering
  \includegraphics[width=0.95\linewidth]{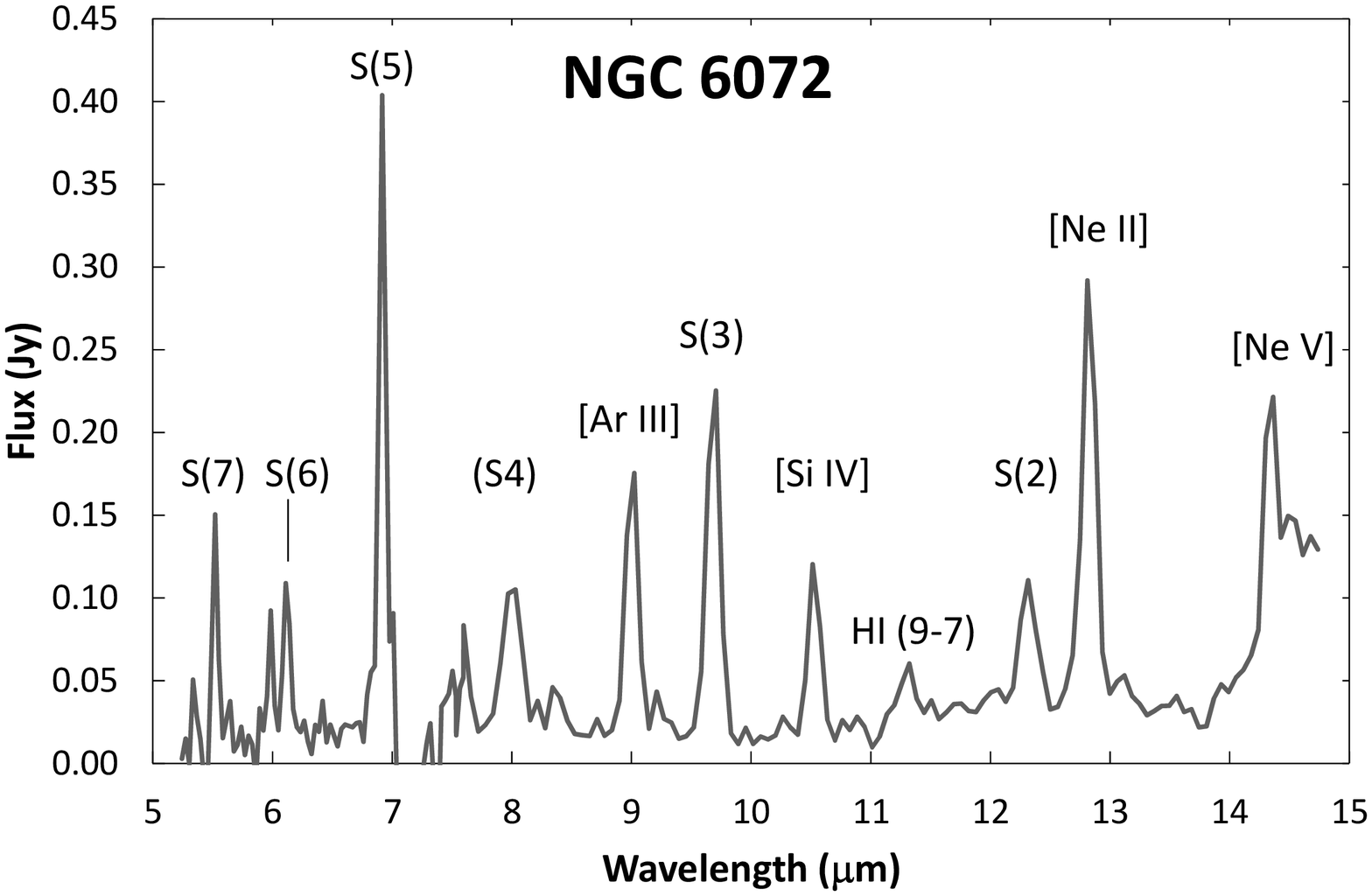}
  \label{fig:NGC6072Flux}
\end{minipage}
\caption{Same as Fig.1 but for NGC\,6072.}
\label{NGC6072Spectra}
\end{figure*}

Figures \ref{M2-51Spectra} to \ref{NGC7293P2Spectra} show optical images 
of each PN obtained from the red band of the Digital Sky Survey (DSS) 
where the location of the on-source and background slit positions are 
marked.  
The one-dimensional mid-IR spectra in the 5.2-14.5 $\mu$m wavelength 
range of each object obtained after sky subtraction and extraction 
are also shown in these figures.  
The 7.4-7.7 $\mu$m spectral range is the overlapping region between SL1 
and SL2 and the reliability in that portion of the spectra is compromised. 
Note the different spectral resolution of the SL1 and SL2 modules: 
emission lines in the 5.2-7.7 $\mu$m SL1 spectral range are narrower 
than those in the 7.5-14.5 $\mu$m SL2 spectral range.

Many important H$_2$ and ionic lines and PAH features are identified 
in these spectra. 

The identification of the broad PAH emission feature at $\simeq$11 
$\mu$m in the SL2 range may be call into question by the possible 
presence of the H~{\sc i} (9-7) transition at 11.3 $\mu$m.  
To assess the nature of the emission feature, when detected, 
we have measured its linewidth and that of nearby ionic or 
H$_2$ molecular lines.  
Only features clearly assymetric and broader than ionic or 
molecular lines have been identified as PAH features. 

Details of the properties of each source and its spectrum are provided 
below.

\subsection{Individual Objects}

\subsubsection{M\,2-51}

\citet{Balick87} classified M\,2-51 as an early elliptical PN.
In the mid-IR, its morphology has also been described as elliptical, with a size of 36$\times$56 arcsec$^{2}$, but with additional extensions along the northwest-southeast direction \citep{Marquez}. The early evolutionary stage, however, has been questioned by \citet{Forveille} and \citet{Saito} who have classified it as an evolved PN.

\begin{figure*}
\vspace{5mm}
\centering
\begin{minipage}{.4\textwidth}
  \centering
  \includegraphics[width=.95\linewidth]{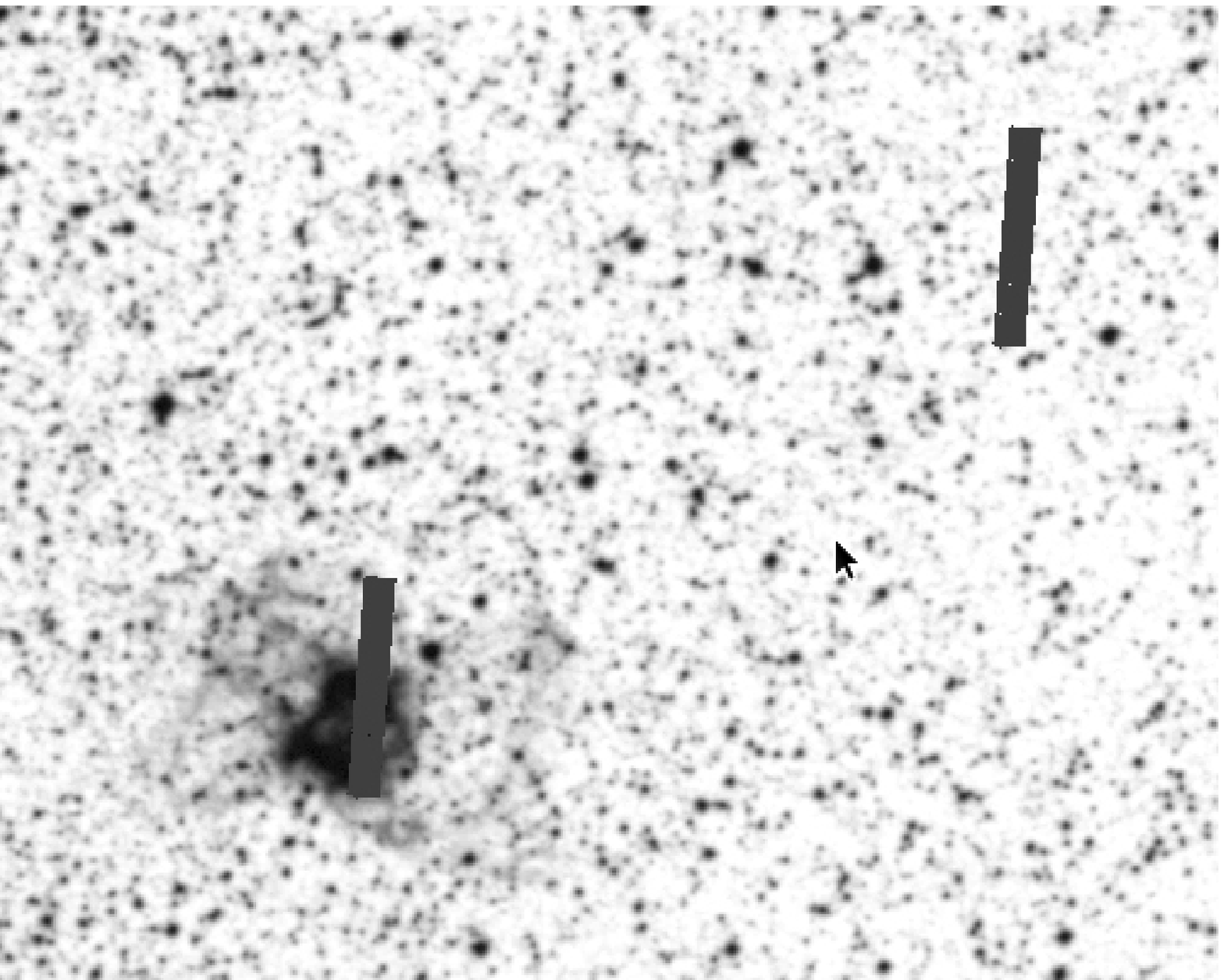}
  \label{fig:NGC6445Slit}
\end{minipage}
\begin{minipage}{.5\textwidth}
  \centering
  \includegraphics[width=0.95\linewidth]{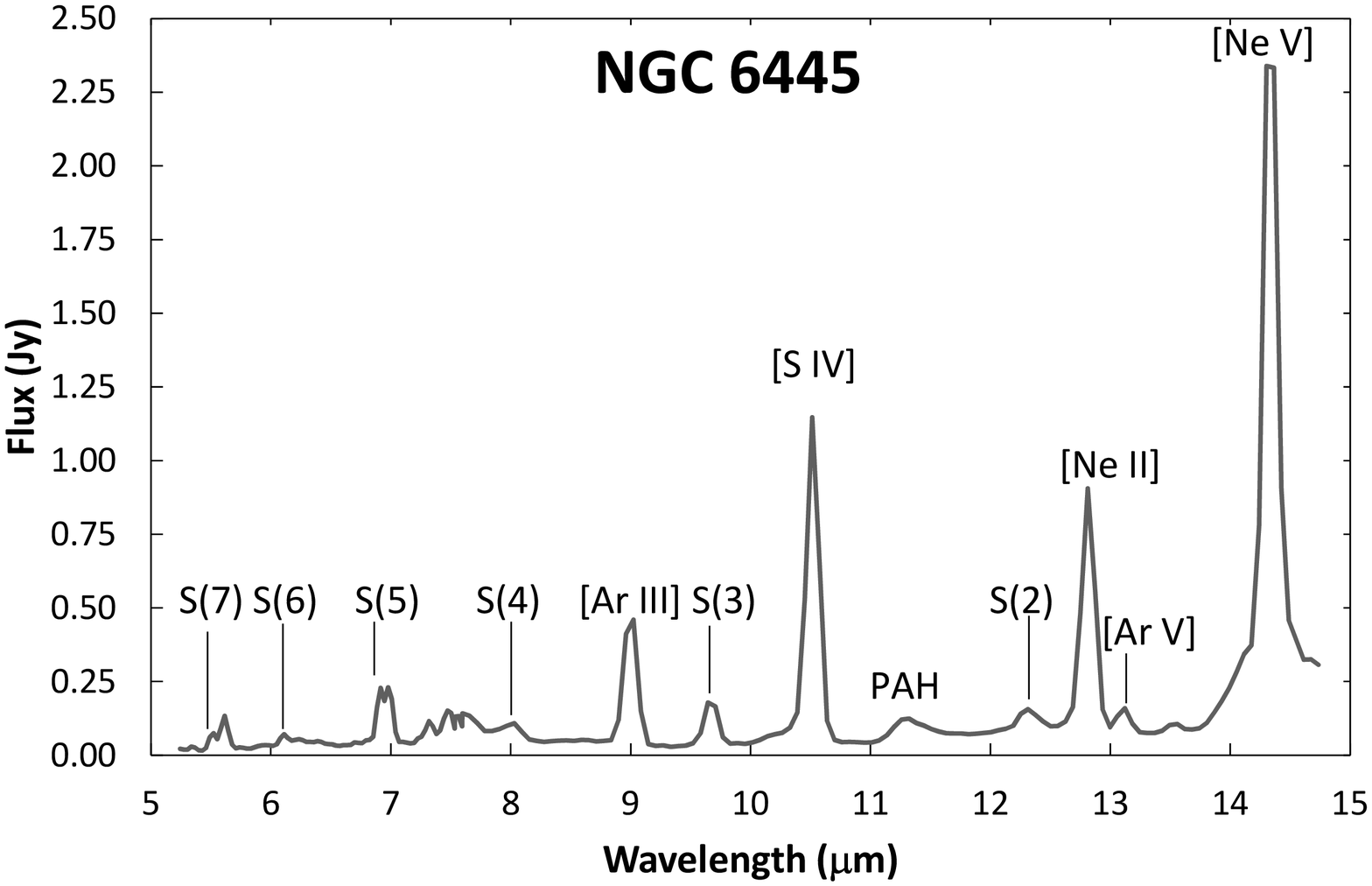}
  \label{fig:NGC6445Flux}
\end{minipage}
\caption{Same as Fig.1 but for NGC\,6445.}
\label{NGC6445Spectra}
\end{figure*}

\begin{figure*}
\vspace{5mm}
\centering
\begin{minipage}{.4\textwidth}
  \centering
  \includegraphics[width=.95\linewidth]{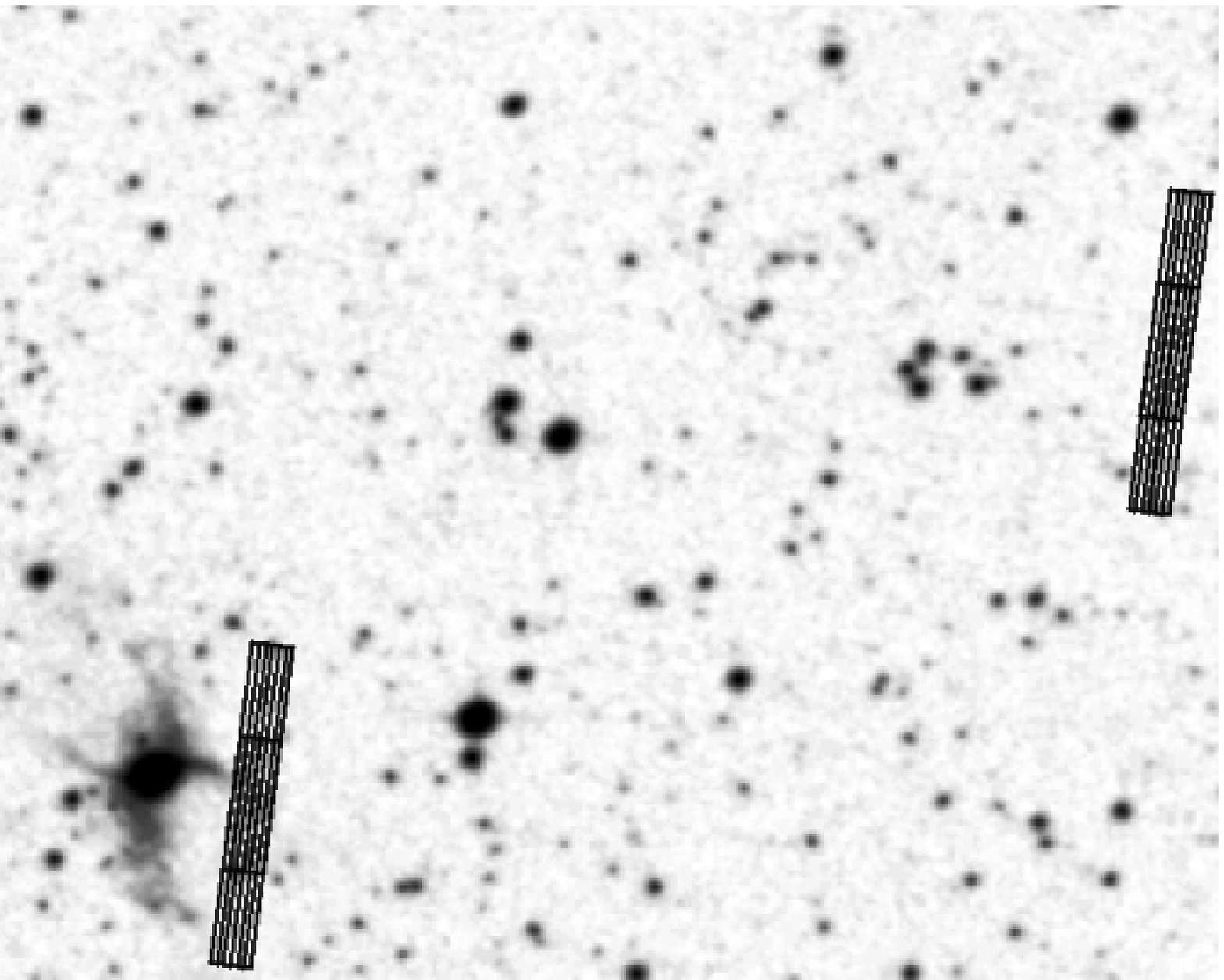}
  \label{fig:NGC6537Slit}
\end{minipage}
\begin{minipage}{.5\textwidth}
  \centering
  \includegraphics[width=0.96\linewidth]{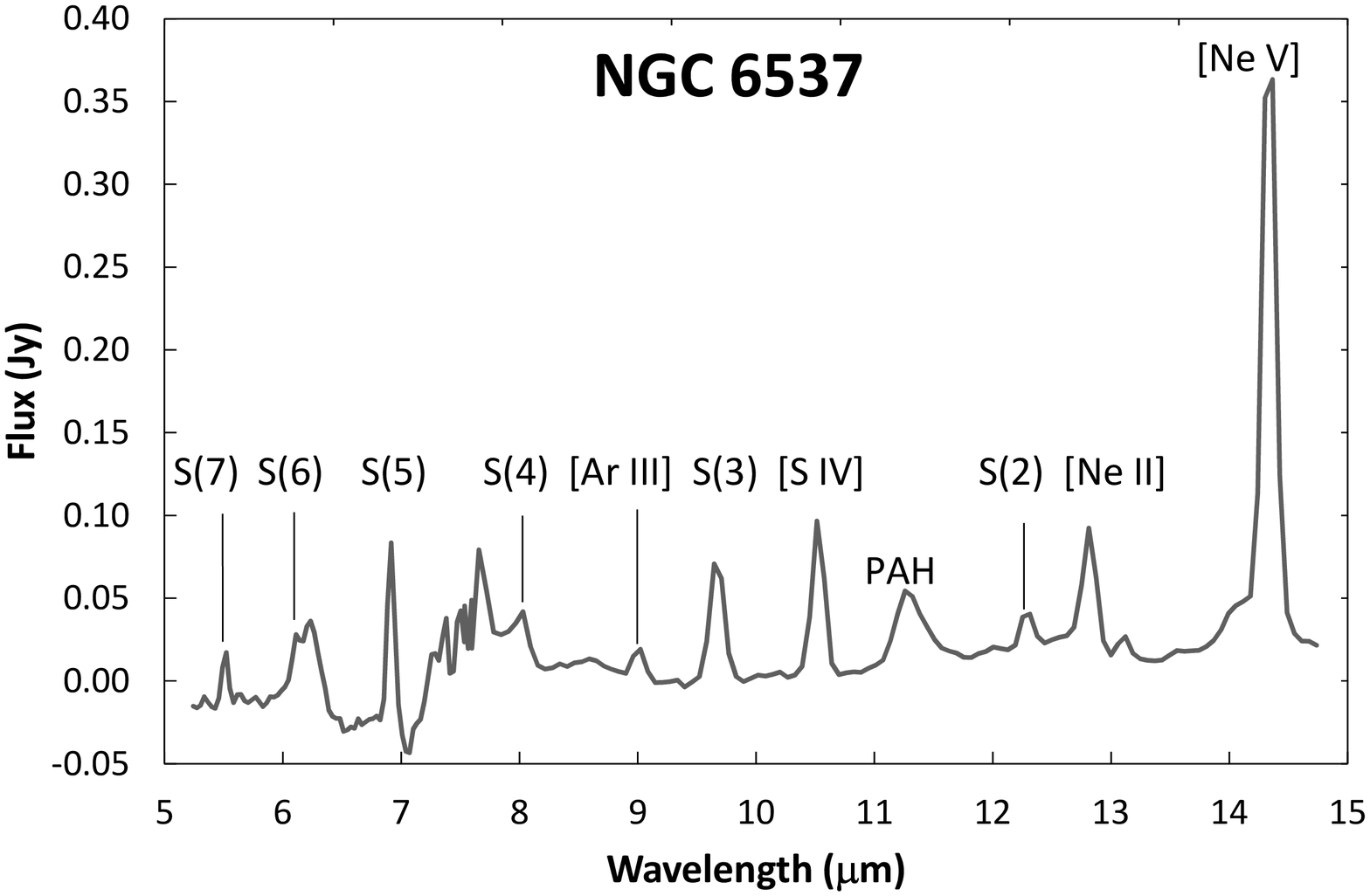}
  \label{fig:NGC6537Flux}
\end{minipage}
\caption{Same as Fig.1 but for NGC\,6537.}
\label{NGC6537Spectra}
\end{figure*}

\begin{figure*}
\vspace{5mm}
\centering
\begin{minipage}{.4\textwidth}
  \centering
  \includegraphics[width=.95\linewidth]{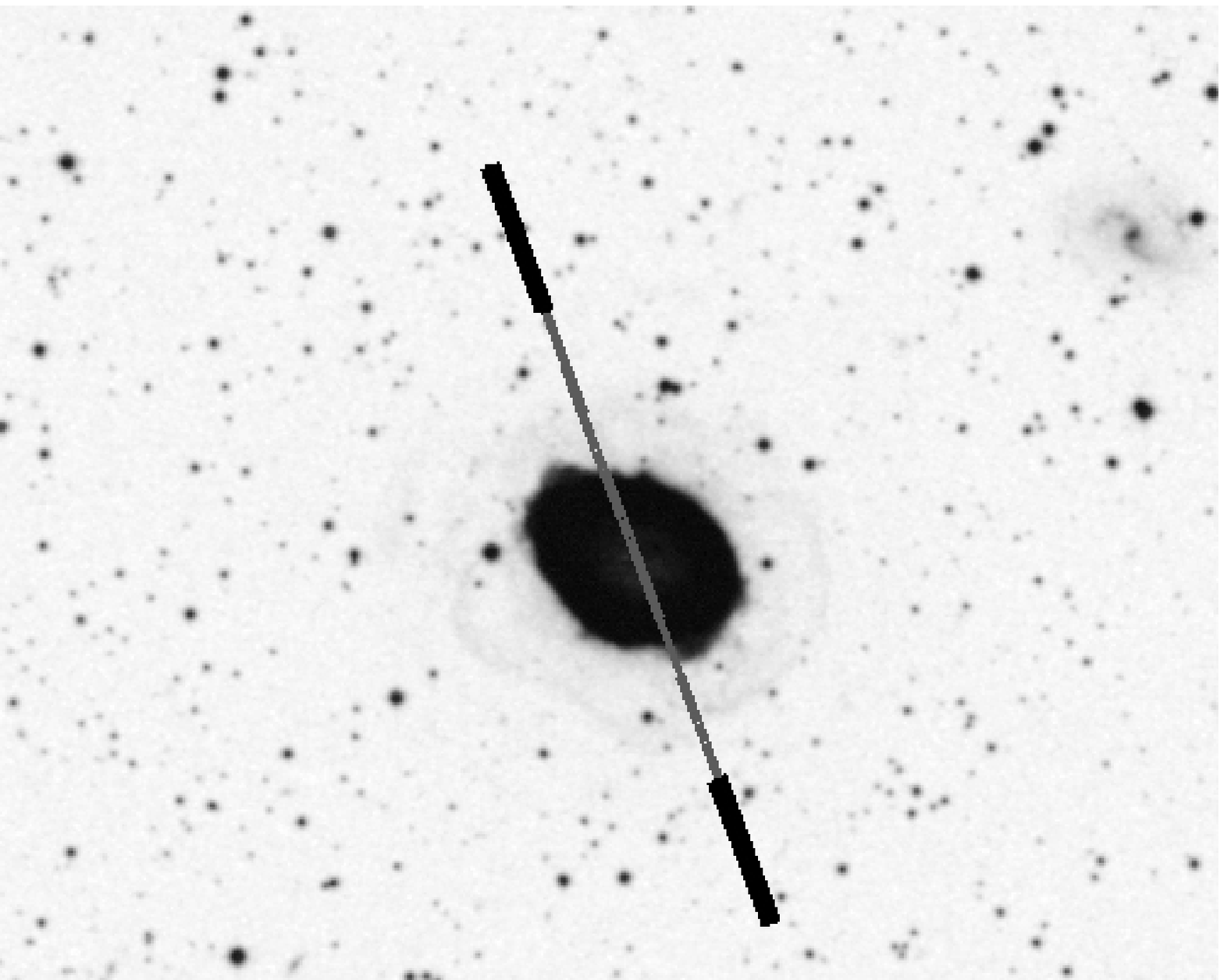}
  \label{fig:NGC6720Slit}
\end{minipage}
\begin{minipage}{.5\textwidth}
  \centering
  \includegraphics[width=0.94\linewidth]{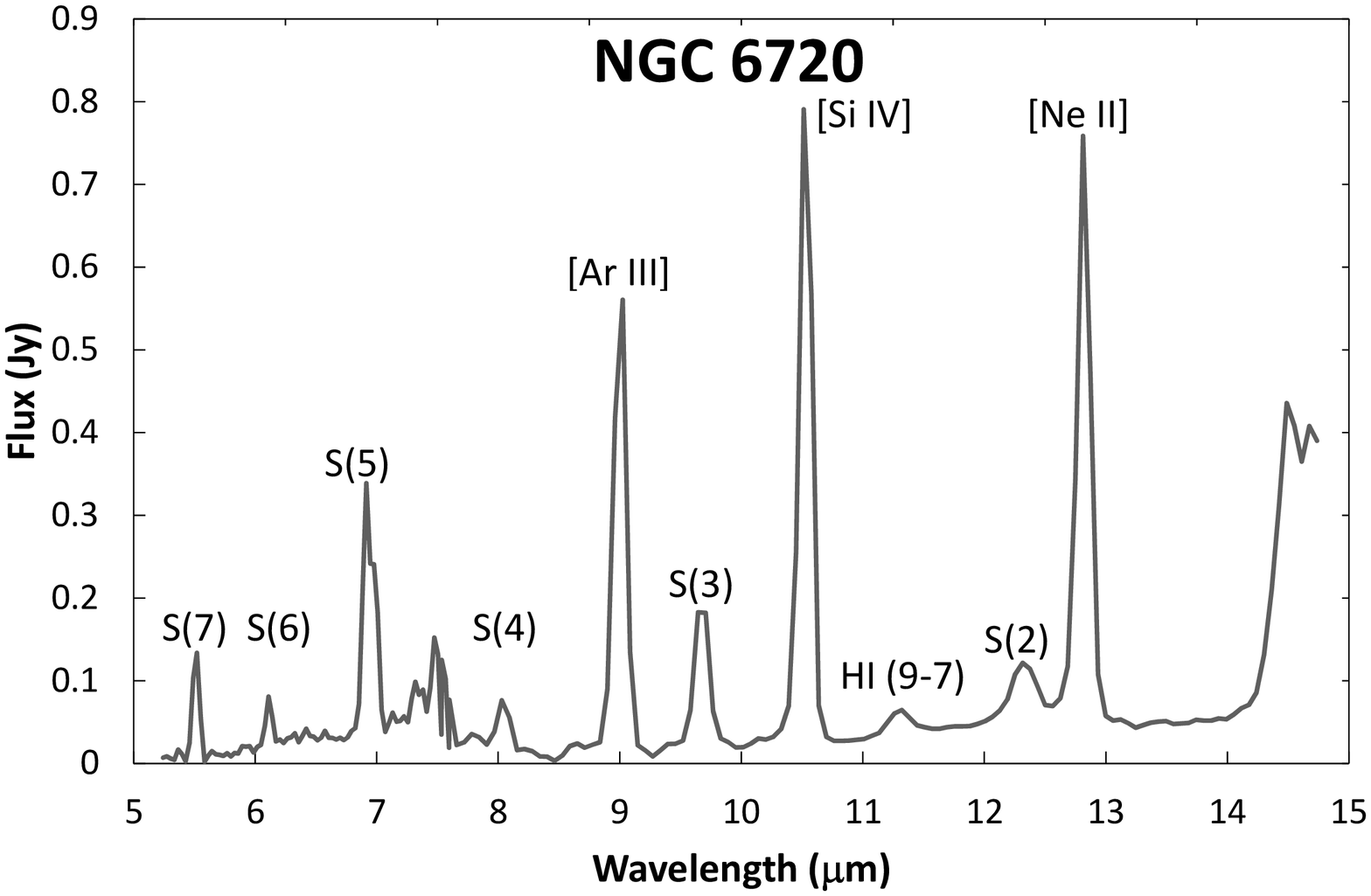}
  \label{fig:NGC6720Flux}
\end{minipage}
\caption{Same as Fig.1 but for NGC\,6720. In this case, the two outer slits in black were used as background.}
\label{NGC6720Spectra}
\end{figure*}

The SL spectrum of M\,2-51 (Figure \ref{M2-51Spectra}) is noisy in the SL1 spectral range, making the identification of molecular hydrogen lines difficult. At any rate, H$_{2}$ lines are present in this spectral region. Two ionic lines were identified in the spectrum, [Ar~{\sc iii}] 
$\lambda$8.99 $\mu$m and [Ne~{\sc ii}] $\lambda$12.81 $\mu$m.  
These spectral features generally agree with those found by \citet{Marquez}, also using the \emph{Spitzer} IRS SL spectra.

\subsubsection{NGC\,2346}

The optical image of NGC 2346 (Figure \ref{NGC2346Spectra}, left) reveals a bipolar morphology with a narrow waist and open bipolar lobes, in accordance with previous descriptions as 
an open-ended butterfly PN. The central star is a binary \citep{Mendez1981} and variable, probably due to dust in orbit around it \citep{Arias}. \citet{Zuckerman} found that the H$_{2}$ emission in the near-IR delineates the equatorial ring and bipolar lobes well, but it is more extended than the emission from ionized gas. The near-IR spectrum is dominated by the H$_{2}$ emission \citep{Hora1999}. Very recently, \citet{Manchado} showed very high resolution imagery in the H$_2$ (1-0) S(1) 2.218 $\mu$m line. For the first time, they were able to resolve the diffuse equatorial ring of NGC\,2346 into an ensemble of molecular knots and filaments, as those reported in other PNe by \citet{Marquez} and references therein.

The mid-IR images of NGC\,2346 show evidence of possible concentric 
rings surrounding the main nebula \citep{Phillips2010}.  
This kind of structure has been observed in several objects 
in the optical and IR (e.g., the Cat's Eye Nebula, NGC\,6543; 
\citet{Balick2001}).  
The mid-IR spectrum (Figure \ref{NGC2346Spectra}) resembles that of 
M\,2-51 in the strength of its lines and noise in the spectral range 
covered by the SL2 module. 
It is worth noting the additional detection of the [S~{\sc iv}] 
$\lambda$10.51 $\mu$m.  
A broad feature at $\simeq$11 $\mu$m might be assigned to 
PAH emission, but the quality of the signal makes this 
identification questionable.

\subsubsection{NGC\,2440}

NGC\,2440 has a complex structure. In shallow optical images (Figure \ref{NGC2440Spectra}, left), it appears as a bipolar PN with its major axis aligned along the east-west direction. Deep exposures reveal up to three distinct bipolar outflows inside the outer main bipolar lobes. The detailed kinematic study presented by \citet{Lopez} definitely confirmed the presence of several bipolar structures emerging along different position angles from the central region. The nebula is bright in H$_2$ in the near-IR spectra  \citep{Hora1999}, however the bipolar structures disappear in near-IR H$_2$ images and an outer, roughly circular halo of H$_{2}$ and radial structures emerge \citep{LatterHora}. In mid-IR images, this round halo shows evidence of interaction with the ISM, with a sharp edge towards the northeast \citep{RamosLarios2009}.

\begin{figure*}
\vspace{5mm}
\centering
\begin{minipage}{.4\textwidth}
  \centering
  \includegraphics[width=.95\linewidth]{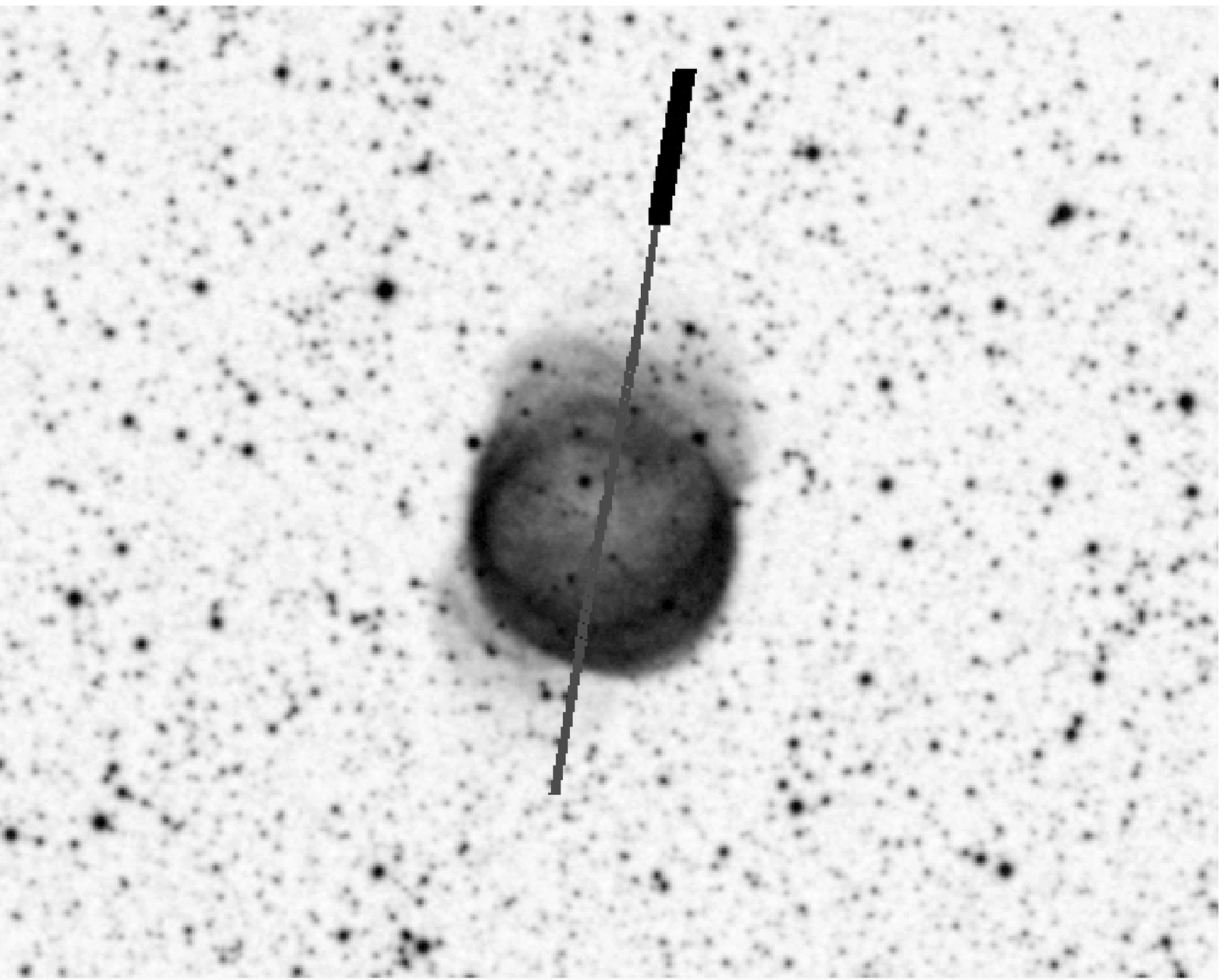}
  \label{fig:NGC6781Slit}
\end{minipage}
\begin{minipage}{.5\textwidth}
  \centering
  \includegraphics[width=0.95\linewidth]{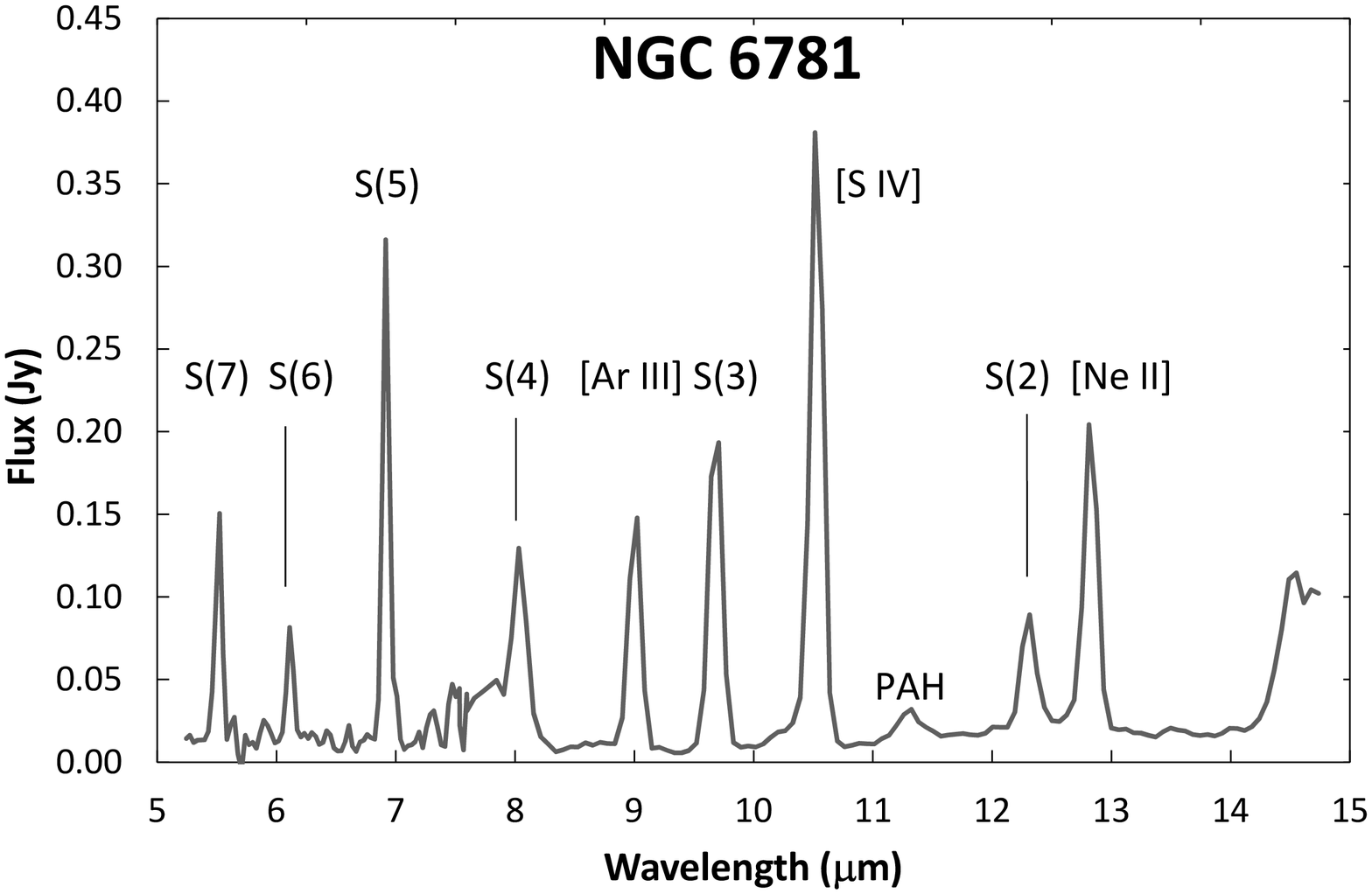}
  \label{fig:NGC6781Flux}
\end{minipage}
\caption{Same as Fig.1 but for NGC\,6781. The upper slit in black was used as background.}
\label{NGC6781Spectra}
\end{figure*}

\begin{figure*}
\vspace{5mm}
\centering
\begin{minipage}{.4\textwidth}
  \centering
  \includegraphics[width=.95\linewidth]{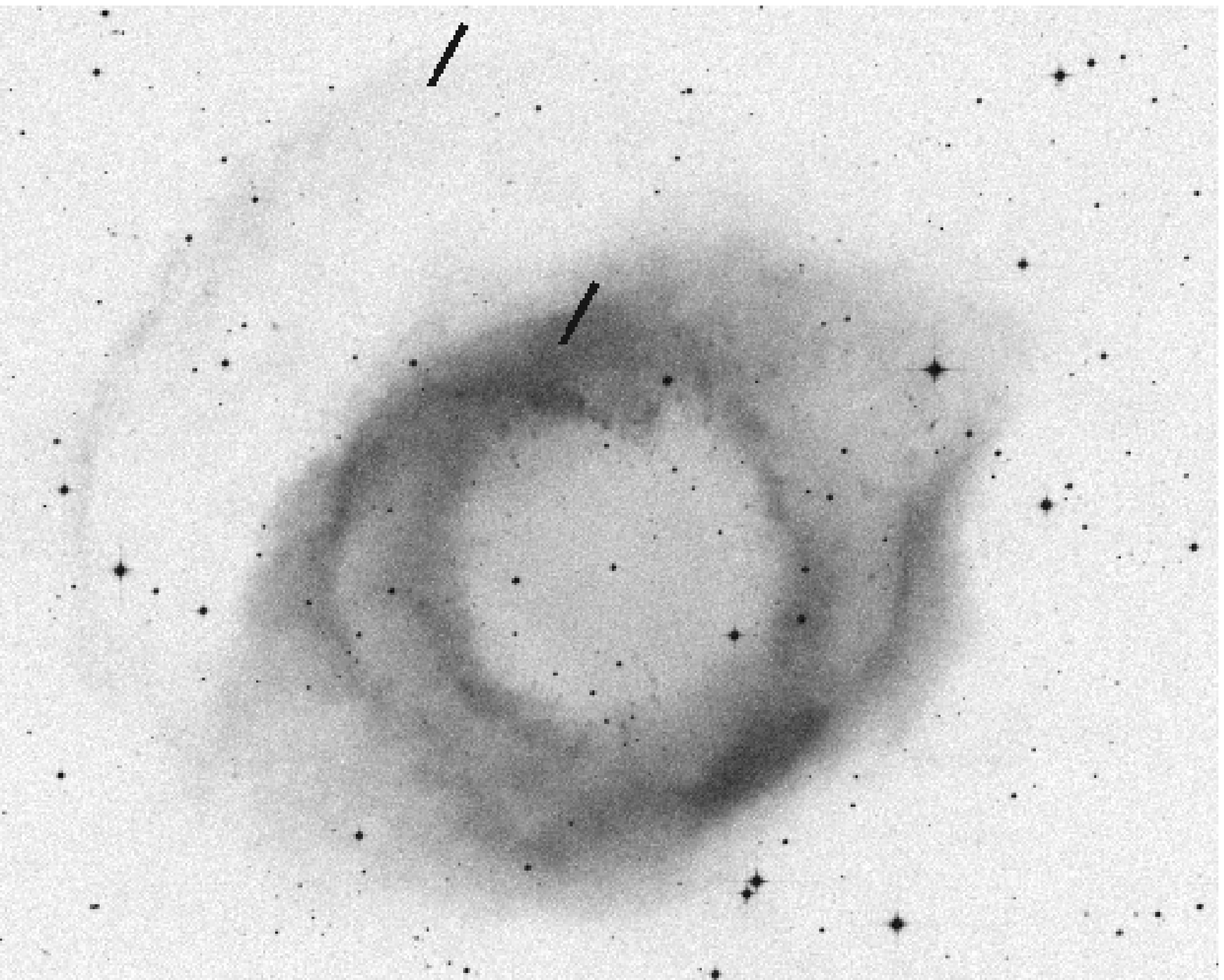}
  \label{fig:NGC7293P1SlitNeg}
\end{minipage}
\begin{minipage}{.5\textwidth}
  \centering
  \includegraphics[width=0.95\linewidth]{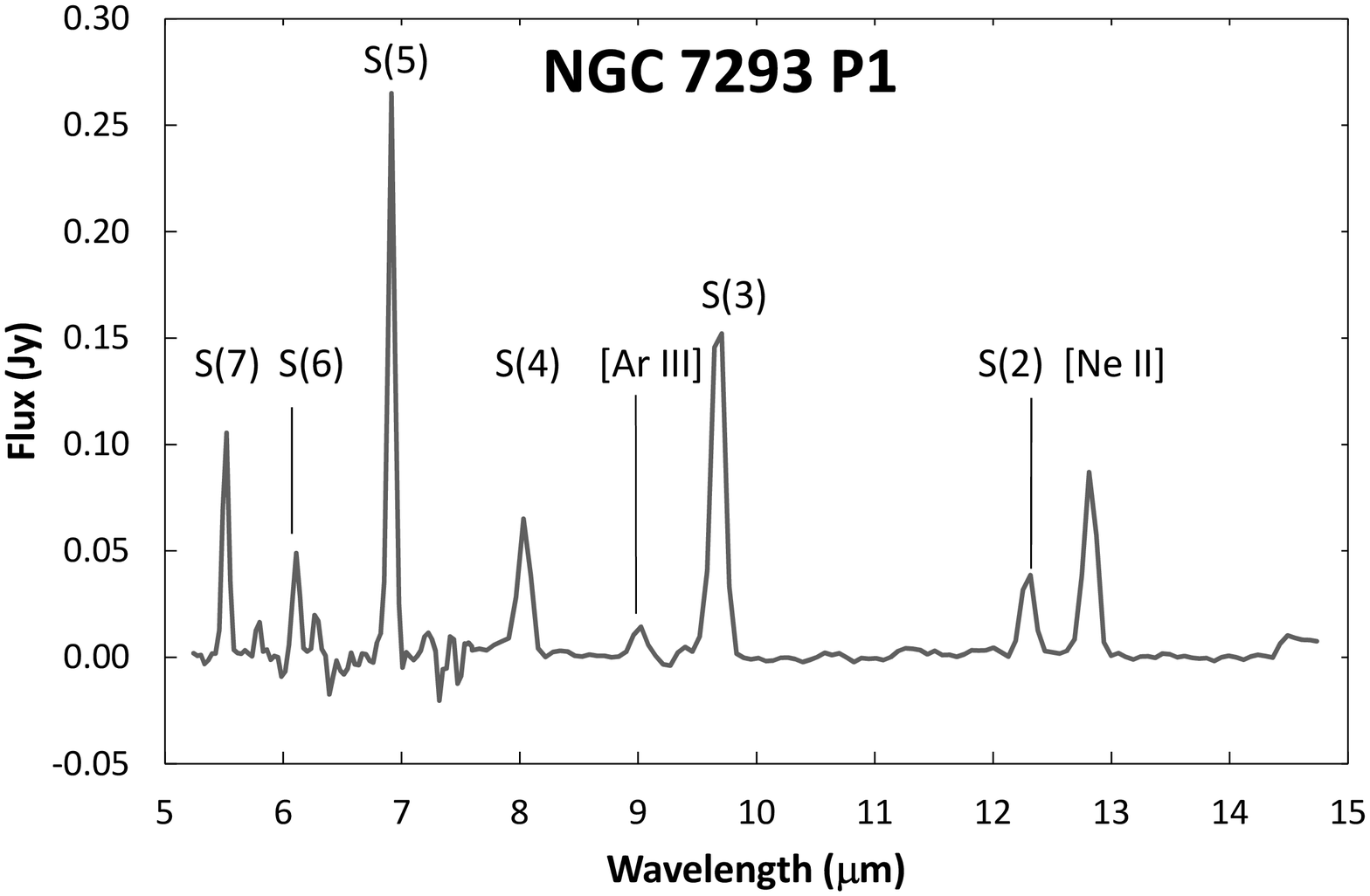}
\end{minipage}
\caption{Same as Fig.1 but for NGC\,7293, position 1. In this case, the upper small slit was used as background.}
\label{NGC7293P1Spectra}
\end{figure*}

\begin{figure*}
\vspace{5mm}
\centering
\begin{minipage}{.4\textwidth}
  \centering
  \includegraphics[width=.95\linewidth]{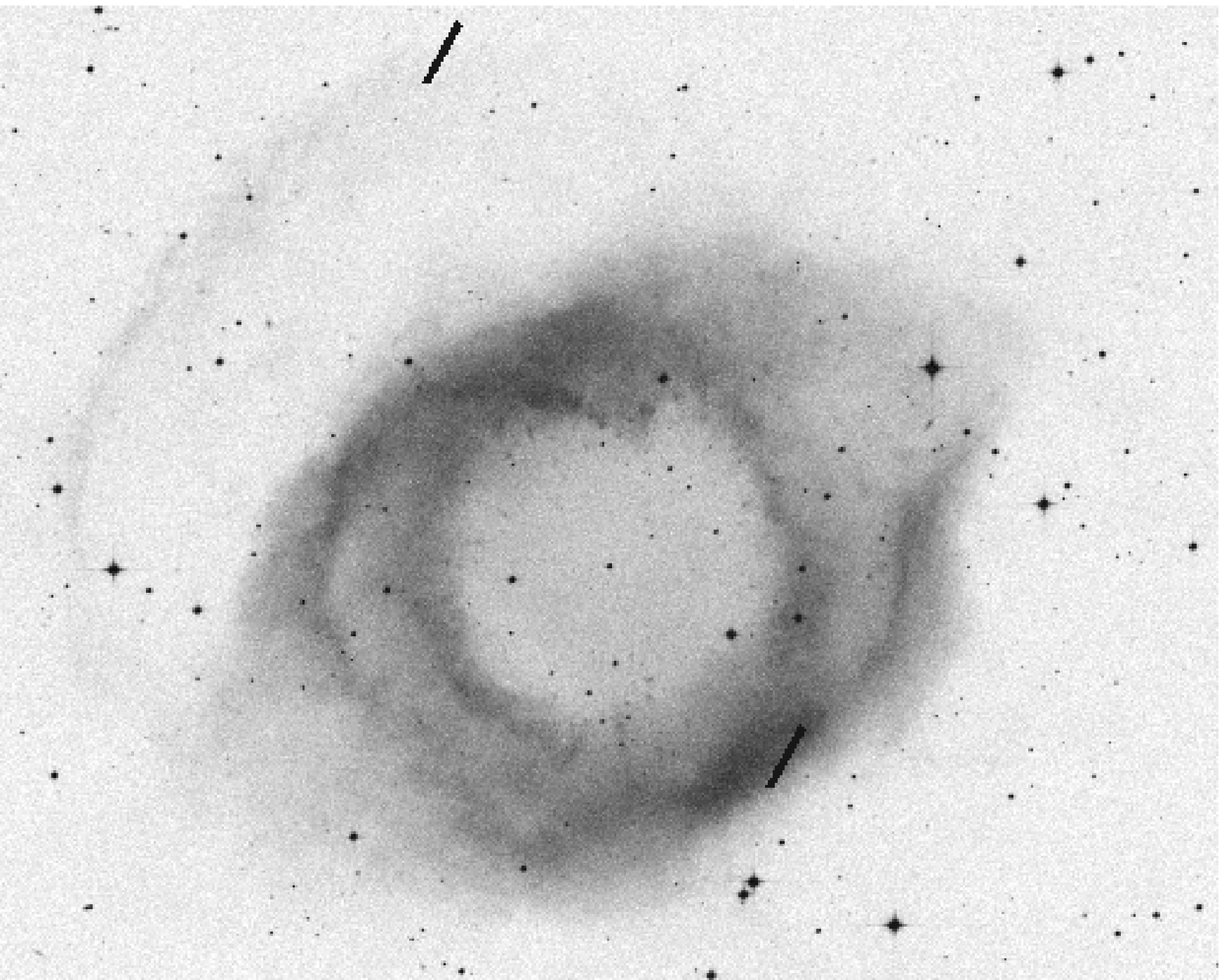}
  \label{fig:NGC7293P2SlitNeg}
\end{minipage}
\begin{minipage}{.5\textwidth}
  \centering
  \includegraphics[width=0.95\linewidth]{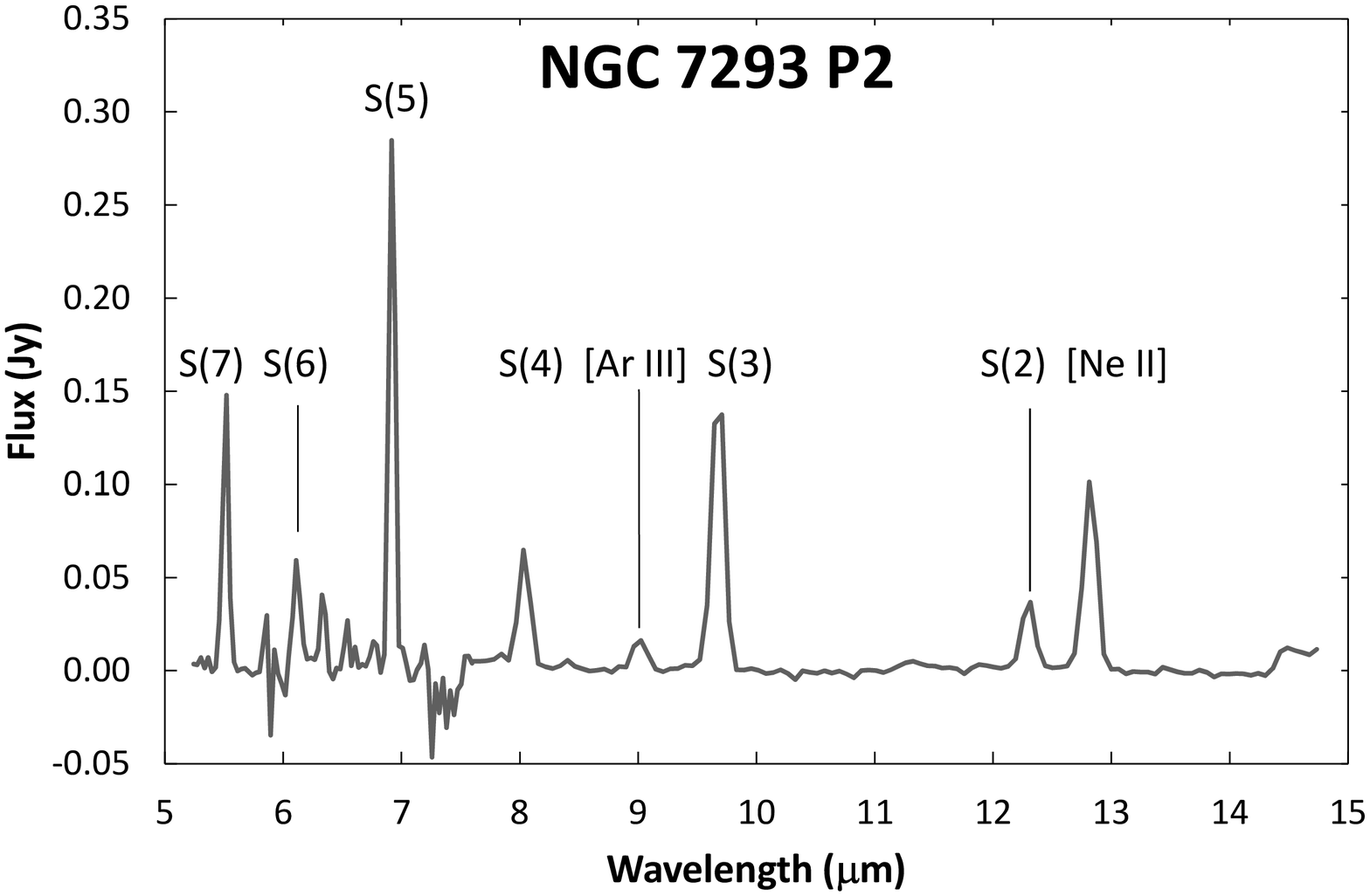}
  \label{fig:NGC7293P2Flux}
\end{minipage}
\caption{Same as Fig.1 but for NGC\,7293, position 2. The upper position was also used as background.}
\label{NGC7293P2Spectra}
\end{figure*}

\begin{figure*}
\vspace{5mm}
\centering
\begin{minipage}{.492\textwidth}
  \centering
  \includegraphics[width=.95\linewidth]{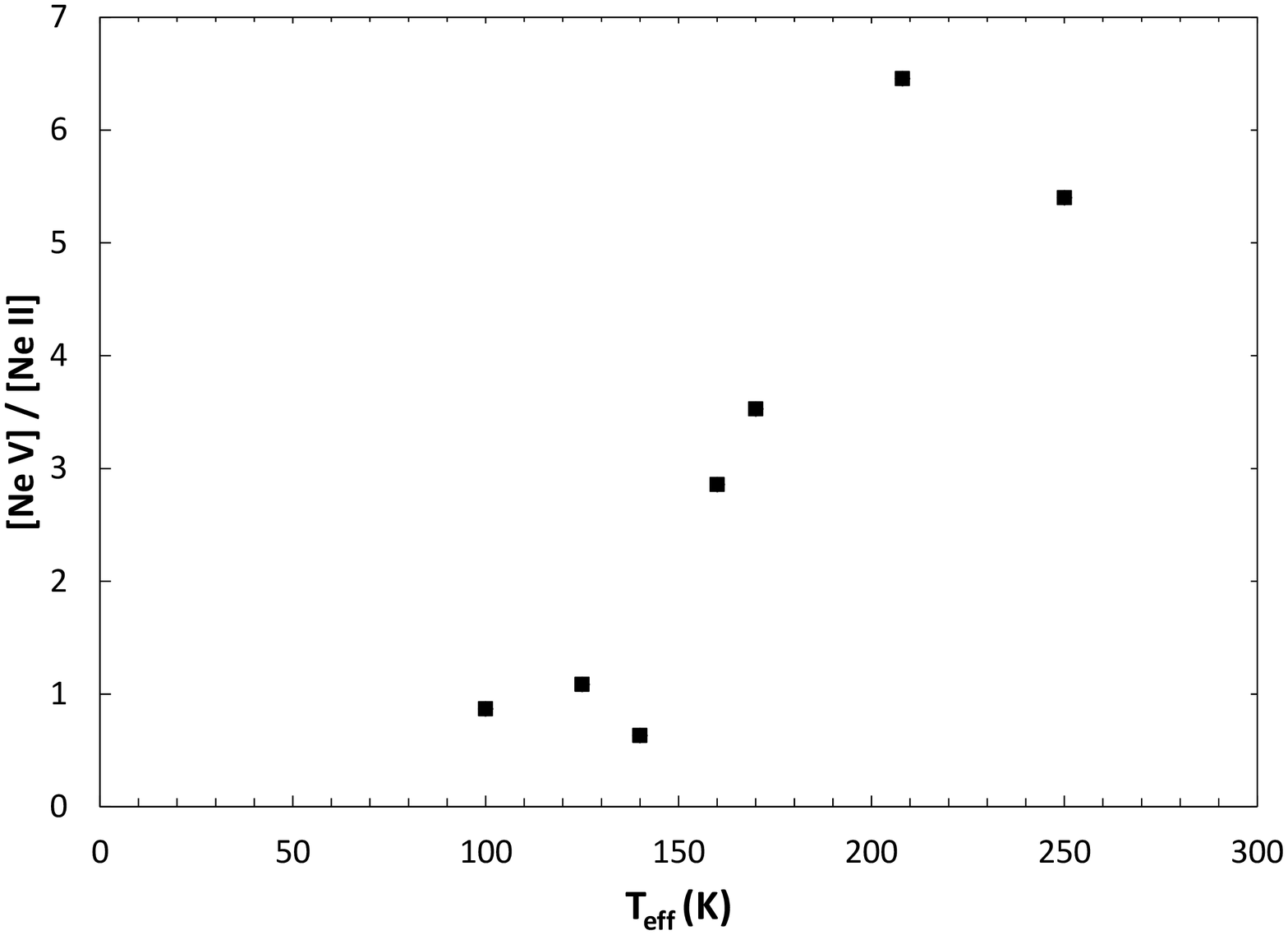}
  \label{fig:NeVNeII}
\end{minipage}
\begin{minipage}{.492\textwidth}
  \centering
  \includegraphics[width=0.95\linewidth]{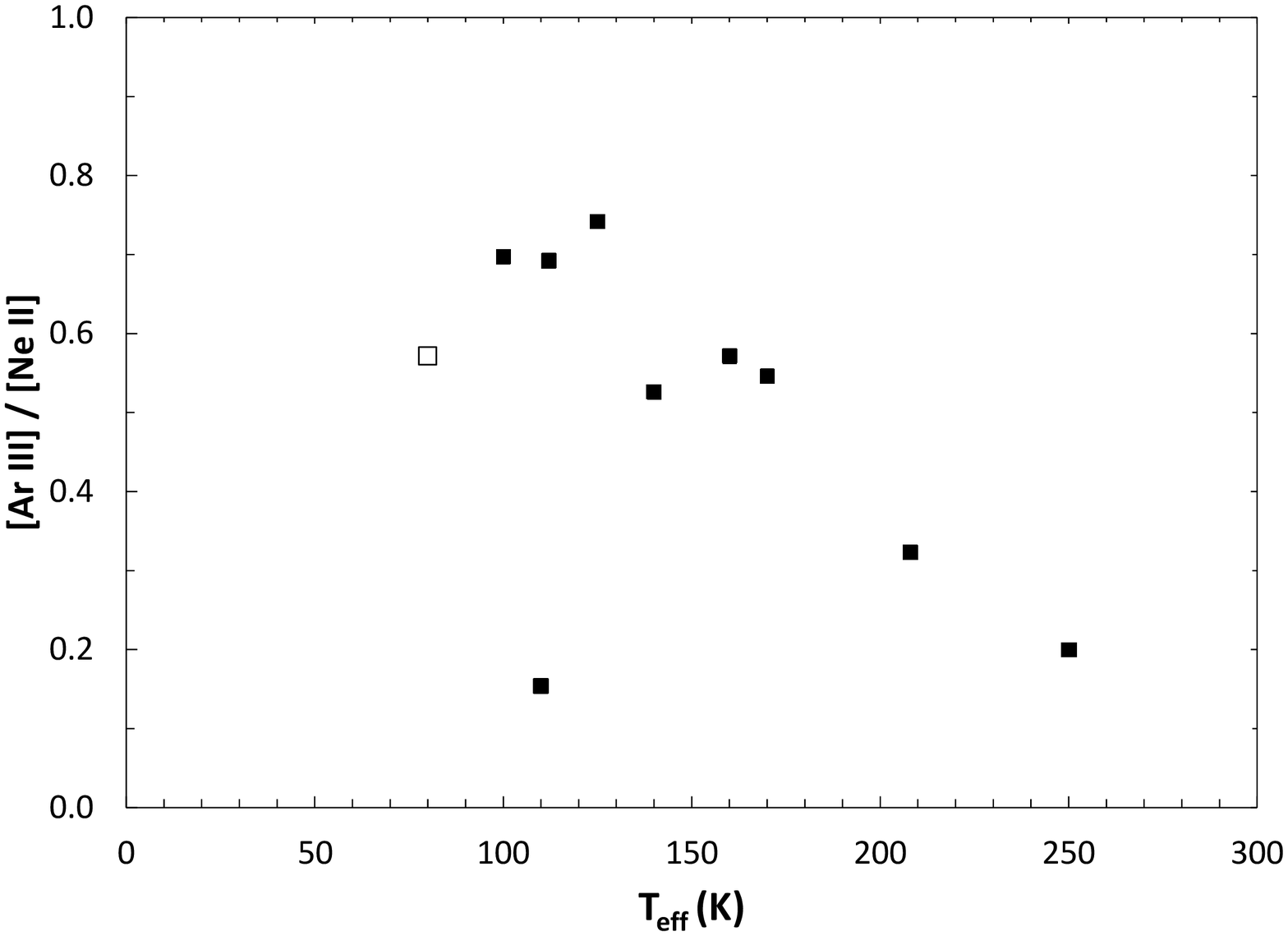}
  \label{fig:ArIIINeII}
\end{minipage}
\caption{
\textit{(left)}[Ne~{\sc v}]/[Ne~{\sc ii}] line ratios for the 
object sample plotted against effective temperature of their 
central stars. 
\textit{(right)}[Ar~{\sc iii}]/[Ne~{\sc ii}] line ratios for 
the sample, also plotted against the central star effective 
temperature. 
The open square corresponds to a lower limit for the effective 
temperature.  
}
\label{LineRatios}
\end{figure*}

The mid-IR spectrum of NGC\,2440 (Figure \ref{NGC2440Spectra}, right) shows contrasting properties to those of the previous sources, M\,2-51 and NGC\,2346. Its spectrum displays a number of weak H$_2$ emission lines also [Ar~{\sc iii}], [S~{\sc iv}], and [Ne~{\sc ii}] lines, but it is dominated by the strong [Ne~{\sc v}] $\lambda$14.32 $\mu$m line. There is a noticeable broad feature at 11.3 $\mu$m, which was identified as the PAH emission. The H$_{2}$ line intensities are generally weaker than those of the ionic species, with S(3) and S(5) being the best-defined.

\subsubsection{NGC\,2818}

The morphology of NGC\,2818 is that of an irregular bipolar.  
The available images in the H$_{2}$ 1-0 S(1) emission line show a sort of elliptical structure delineating the H$\alpha$ emission from the equatorial waist and the bipolar lobes \citep{Schild,Bohigasa}. In addition, the excellent colour-composite three-colour image in the \textit{Spitzer} Infrared Array Camera (IRAC) 3.6, 4.5 and 8.0 $\mu$m bands of \citet{Hora2007} reveals faint extended emission that closely matches that of the H$_{2}$ in the near-IR. Assuming a homogeneous expansion rate, the expansion velocity of NGC\,2818 implies a kinematic age $\sim$8400 years \citep{Vazquez}.

The mid-IR spectrum of NGC\,2818 (Figure \ref{NGC2818Spectra}-{\it right}) 
shows very well-defined, bright emission lines for all the H$_{2}$ lines in 
the SL1-SL2 spectral range. 
The S(5) and S(3) lines are the most prominent among the 
molecular hydrogen transitions. 
Ionic lines of [Ar~{\sc iii}], [S~{\sc iv}], [Ne~{\sc ii}] and 
[Ne~{\sc v}] are also detected, as well as a weak
emission line at $\sim$11.3 $\mu$m that can be identified with 
the H~{\sc i} (9-7) recombination line.

\subsubsection{NGC\,3132}

Despite its ellipsoidal morphology, \citet{Monteiro} found that a bipolar ``Diabolo'' represented the physical structure of NGC\,3132 more accurately.  This PN is one of the H$_2$ brightest \citep{Storey1984}, with an H$_{2}$ spatial distribution following the optical morphology. The H$_2$ line ratios derived from near-IR spectra implies shock excitation \citep{Zuckerman}.

The \emph{Spitzer} mid-IR IRS spectrum shows all previously identified H$_{2}$ and ionic emission lines (Figure~\ref{NGC3132Spectra}).  The ionic emission lines of [Ar~{\sc iii}], [S~{\sc iv}], and [Ne~{\sc ii}] are the brightest in the spectrum.

\subsubsection{NGC\,6072}

NGC\,6072 has the appearance of an elliptical ring along the north-south direction, with faint extensions in the east-west direction.  The nature of this ring is emphasized by its strong CO emission and expansion velocity of the molecular material, $\sim$15 km~s$^{-1}$ \citep{Cox1991}. The detailed analysis of near-IR H$_{2}$ and \textit{Spitzer} IRAC images of NGC\,6072 presented by \citet{Kwok6072} reveals a quadrupolar morphology \citep{Manchado1996}, with a prominent equatorial ring and hints of additional multipolar features. Analyses of near-IR spectra conclude that the H$_{2}$ is primarily excited by shocks \citep{Zuckerman}.  

The mid-IR spectrum of NGC\,6072 is somewhat noisier than the previous 
ones (Figure \ref{NGC6072Spectra}-{\it right}).
The very same transitions found in the mid-IR spectrum of NGC\,3132 were 
found for NGC\,6072.  
The brightest lines in the mid-IR spectrum according to our measurements 
are H$_2$ 0-0 S(5) and [Ne~{\sc ii}].

\subsubsection{NGC\,6445}

NGC\,6445 is characterized by an inner square-shaped ring surrounded by a faint intricate bipolar structure \citep{CuestaPhillips}. 
Based on observations of its density, the nebula appears to be evolved.  
The mid-IR \textit{Spitzer} image by \citet{Phillips2010} shows differences in the optical and mid-IR distributions, and the presence of strong 8.0 $\mu$m emission outside the central nebula. They argue the possibility that this extended mid-IR emission derives from a cylinder of neutral material that would be responsible for the large-scale collimation of the optical structures.  

The mid-IR spectrum of NGC\,6445 (Figure \ref{NGC6445Spectra}-{\it right}) resembles closely that of NGC\,2440, although with relatively brighter emission in the [S~{\sc iv}] and [Ne~{\sc ii}] lines compared to that of [Ne~{\sc v}].  
The H$_2$ transition lines are a bit brighter in NGC\,6445 than in NGC\,2440, though still fainter than the ionic lines.

\subsubsection{NGC\,6537}

NGC\,6537, also known as the Red Spider Nebula, has a pronounced bipolar shape.  The central star is located behind a dense shell of dust, making its properties difficult to determine.  Even so, its temperature is estimated to be at least 180,000 K \citep{Hyung}.

The spectrum obtained for NGC\,6537 (Figure \ref{NGC6537Spectra}-{\it right}) is largely dominated by the [Ne~{\sc v}] $\lambda$14.32 $\mu$m line at the right end of the spectrum.  Other emission lines are also easily identifiable.  The [Ar~{\sc iii}] line is noticeably less prominent than in other spectra, and H$_2$ transitions are present.  The spectrum obtained in this study is similar to the one presented by \citet{Marquez} for the lobular region of the nebula.

\begin{table*}
\tabcolsep=2.5pt
\centering
\caption{Line fluxes ($\times$ 10$^{-13}$ erg~s$^{-1}$~cm$^{-2}$~\AA$^{-1}$)}
\label{FluxSummary}
\begin{tabular}{@{}lcccccccccccccc@{}}
\toprule
Transition  & 
H$_2$ S(7)  & 
H$_2$ S(6)  & 
H$_2$ S(5)  & 
{[}Ar \textsc{ii}{]} & 
H$_2$ S(4) & 
{[}Ar \textsc{iii}{]} & 
H$_2$ S(3) & 
{[}S \textsc{iv}{]} & 
H\textsc{i} (9-7) & 
H$_2$ S(2) & 
{[}Ne \textsc{ii}{]} & 
{[}Ar \textsc{v}{]} & 
{[}Ne \textsc{v}{]} &
PAH\\ 
$\lambda$ ($\mu$m)      & 5.51    & 6.11    & 6.91    & 6.99        & 8.02    & 8.99         & 9.66    & 10.51      & 11.3   & 12.28     & 12.81       & 13.1       & 14.32   &    \\ 
Source & & & & & & & & & & & & & & \\  \midrule
M\,2-51      &  8 & 4 & 14 & $\cdots$ & 11 &  6 & 20 & $\cdots$ & $\cdots$ & 8  &  13 & $\cdots$ & $\cdots$ & no  \\
NGC\,2346    & 13 & 2 & 23 & $\cdots$ &  7 &  8 & 21 & 7        & $\cdots$ & 7  &  14 & $\cdots$ & $\cdots$ & ?  \\
NGC\,2440    &  3 & 3 &  9 & $\cdots$ &  8 & 22 & 12 & 58       & $\cdots$ & 7  &  68 & 10       & 439      & yes \\
NGC\,2818    &  3 & 2 & 10 & $\cdots$ &  4 &  4 & 13 & 22       & 0.1      & 5  &   7 & 1        &  20      & no  \\
NGC\,3132    & 11 & 8 & 38 & $\cdots$ & 11 & 53 & 35 & 83       & $\cdots$ & 14 &  76 & $\cdots$ &  66      & yes \\
NGC\,6072    & 10 & 6 & 30 & $\cdots$ & 15 & 20 & 30 & 13       &  5       & 13 &  38 & $\cdots$ &  24      & no  \\
NGC\,6445    &  4 & 3 & 12 & 13       &  6 & 59 & 19 & 137      & $\cdots$ & 12 & 108 & 9        & 381      & yes \\
NGC\,6537    &  2 & 3 &  8 & $\cdots$ &  6 &  2 & 11 & 12       & $\cdots$ & 3  &  10 & $\cdots$ & 54       & yes \\
NGC\,6720    & 10 & 3 & 20 & $\cdots$ &  7 & 69 & 25 & 99       &  8       & 14 &  93 & $\cdots$ & 101      & no \\
NGC\,6781    &  9 & 4 & 21 & $\cdots$ & 11 & 18 & 27 & 50       & $\cdots$ & 10 &  26 & $\cdots$ & $\cdots$ & yes \\
NGC\,7293 P1 &  7 & 3 & 19 & $\cdots$ &  8 &  2 & 23 & $\cdots$ & $\cdots$ & 6  &  12 & $\cdots$ & $\cdots$ & no\\
NGC\,7293 P2 &  9 & 4 & 18 & $\cdots$ &  7 &  2 & 20 & $\cdots$ & $\cdots$ & 5  &  14 & $\cdots$ & $\cdots$ & no\\ \bottomrule
\end{tabular}
\end{table*}

\subsubsection{NGC\,6720}

The well studied Ring Nebula (NGC\,6720 or M\,57) is an ellipsoidal shell with a bright equatorial region seen almost pole-on \citep{Guerrero1997,ODell07}.  The expansion velocity of the nebula is 20-30 km\,s$^{-1}$.  
The surface temperature of its central star is 125,000 K \citep{vanHoof2010}.

The near-IR spectrum of the central ring shows prominent H$_2$ emission lines \citep{Hora1999}.  The mid-IR spectrum, on the other hand, is dominated by the ionic lines of [S~{\sc iv}], [Ne~{\sc ii}] and [Ar~{\sc iii}], with weaker H$_2$ 
transitions (Figure \ref{NGC6720Spectra}-{\it right}).  
The S(5) and S(3) are the most prominent H$_2$ lines.

\subsubsection{NGC\,6781}

Even though NGC 6781 presents an elliptical morphology, it is very likely a bipolar nebula seen almost pole-on. The H$_{2}$ envelope forms a thin hollow cylinder which is open at both ends and tilted towards the line of sight, with an expansion velocity of 22 km\,s$^{-1}$ \citep{Hiriart}.
It shares with other PNe the peculiar property that the expansion velocity of its molecular gas is greater than the expansion velocity of the ionized gas.  An optical and infrared study of this object by \citet{Phillips2011} found 
an H$_{2}$ excitation temperature of $\sim$980 K.

Ionic transitions of [S~{\sc iv}], [Ar~{\sc iii}], and 
[Ne~{\sc ii}] are found in the mid-IR spectrum of NGC 6781 
(Figure \ref{NGC6781Spectra}-{\it right}).  Molecular hydrogen lines are also present, but they are weaker than 
the [S~{\sc iv}] emission line.

\begin{figure*}
\vspace{3mm}
    \centering
    \subfloat[]{
        \label{fig:volt1cnv}
        \includegraphics[width=0.482\linewidth]{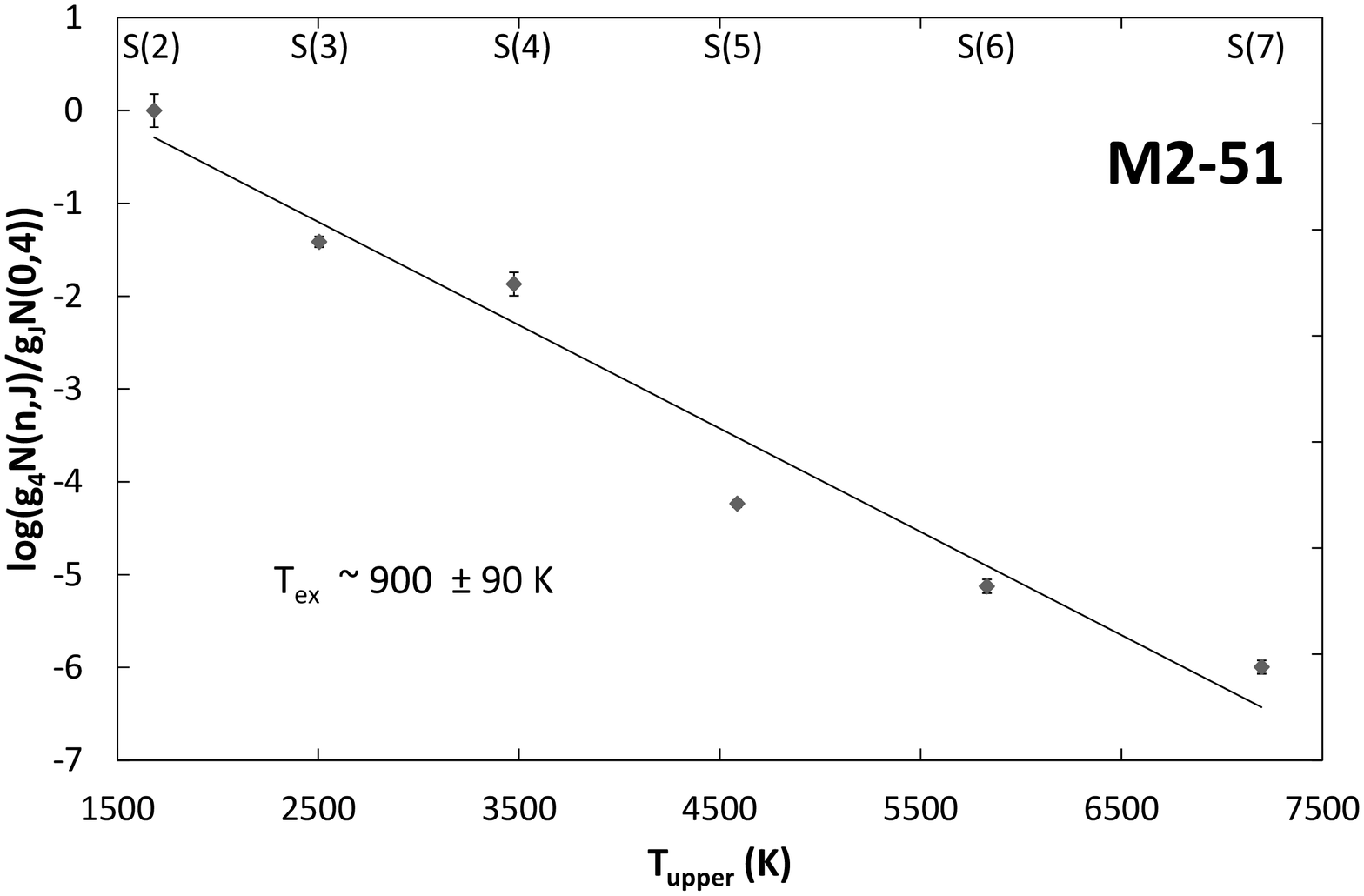}}
    \hspace{0.02\linewidth}
    \subfloat[]{
        \label{fig:crt1cnv}
        \includegraphics[width=0.485\linewidth]{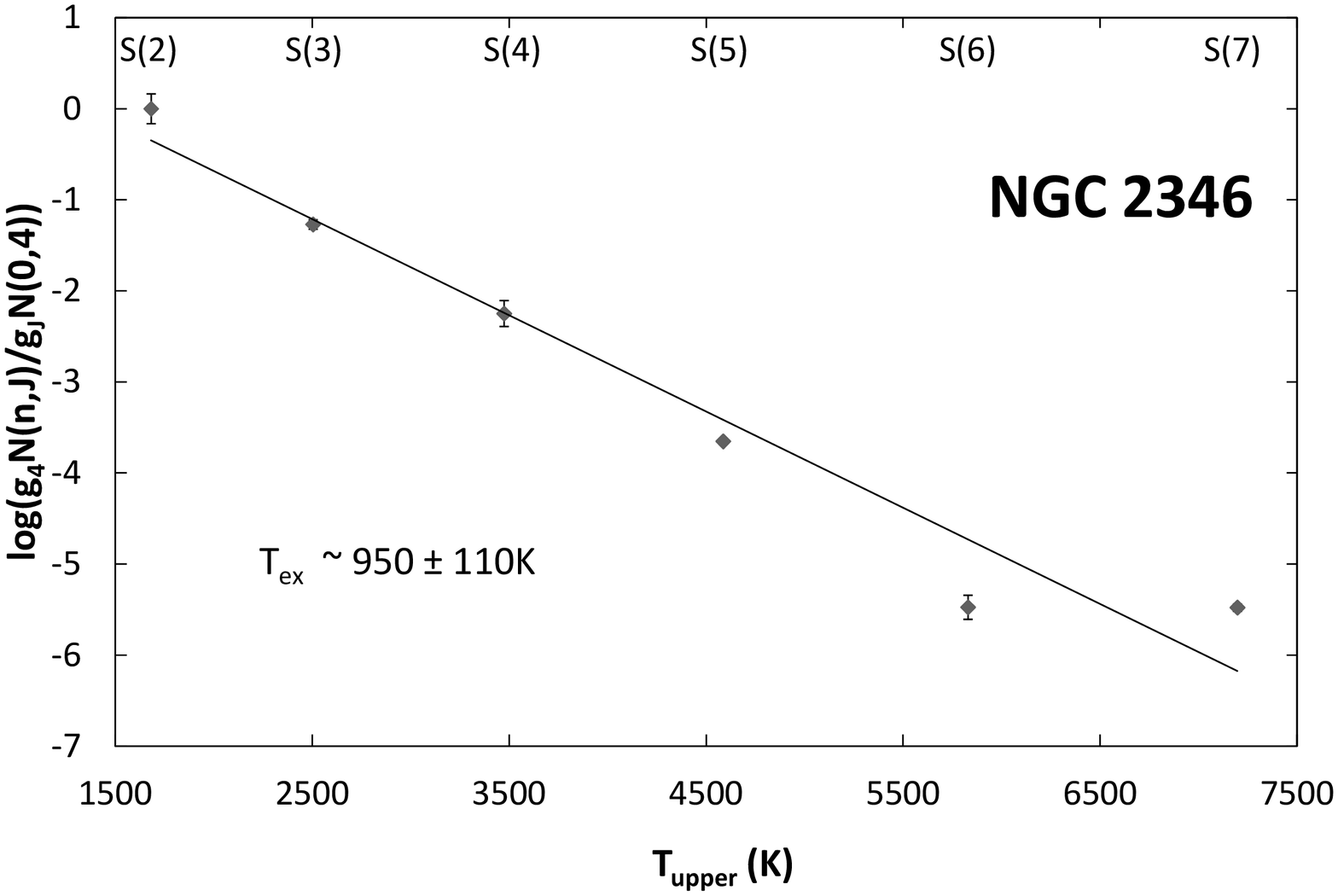}}\\
    \subfloat[]{
        \label{fig:volt1cnv}
        \includegraphics[width=0.468\linewidth]{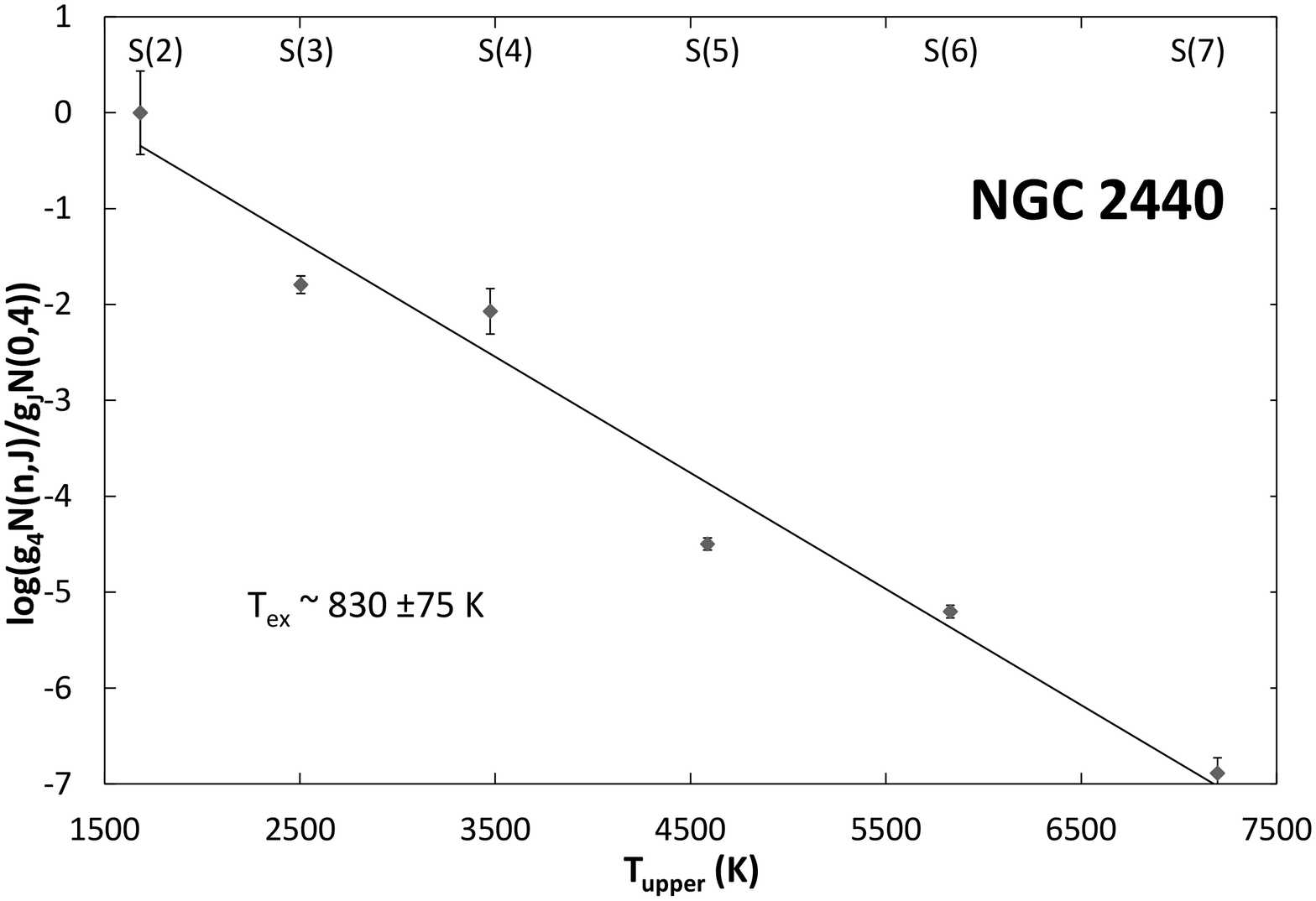}}
    \hspace{0.02\linewidth}
    \subfloat[]{
        \label{fig:crt1cnv}
        \includegraphics[width=0.482\linewidth]{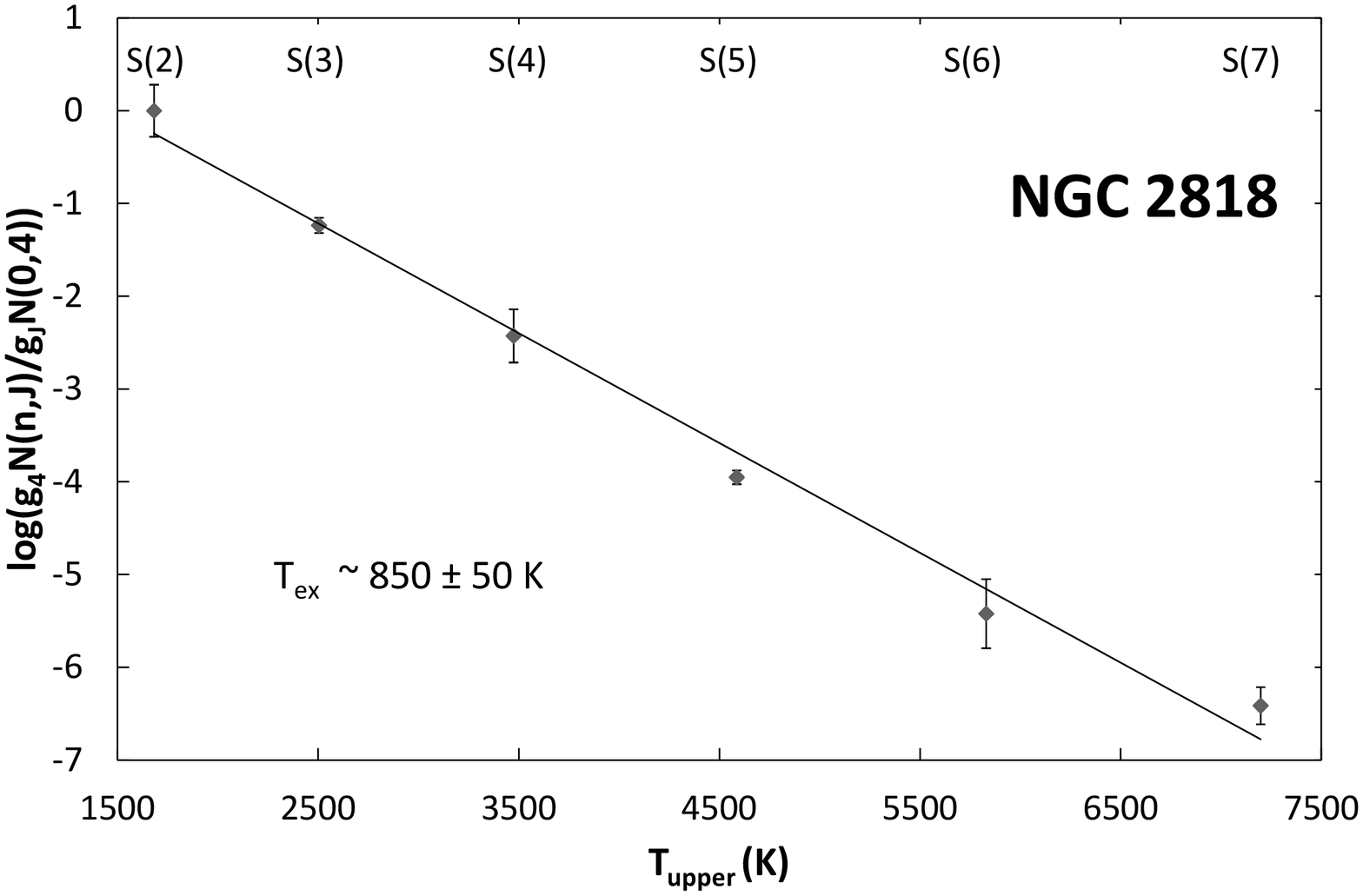}}\\
    \subfloat[]{
        \label{fig:volt1cnv}
        \includegraphics[width=0.482\linewidth]{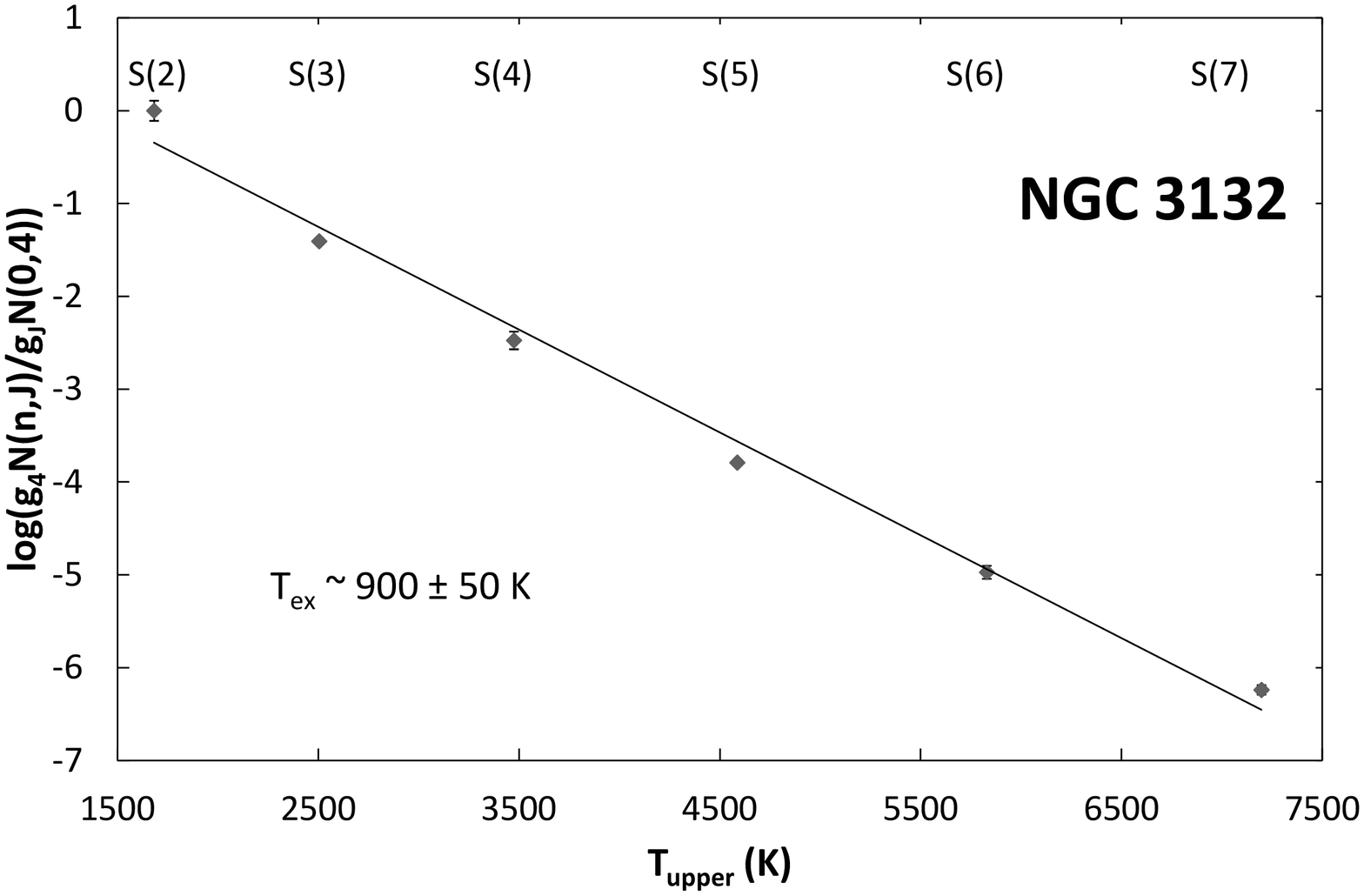}}
    \hspace{0.02\linewidth}
    \subfloat[]{
        \label{fig:crt1cnv}
        \includegraphics[width=0.482\linewidth]{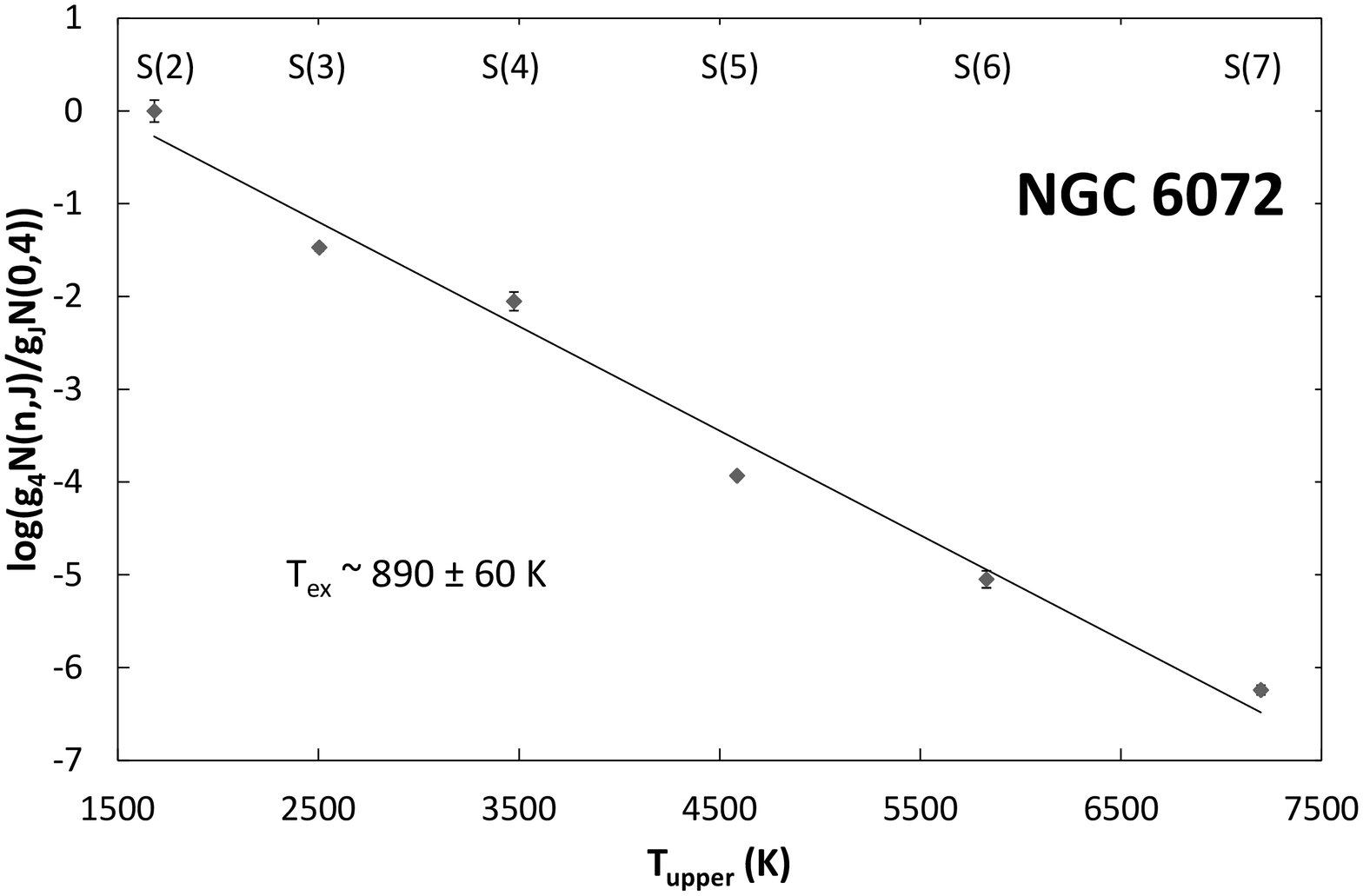}}\\
            \caption{
The mean variation in H$_{2}$ populations for
the $\nu$ = 0-0 S(2)-S(7) transitions is represented in the excitation diagrams for M\,2$-$51, NGC\,2346, NGC\,2440, NGC\,2818, NGC\,3132 and NGC\,6072. The results are based on averages of line strengths for spectral selected positions (error bars are included).
The least-square fit to the results implies rotational excitation temperatures
between $\sim$ 800 - 950 K.}
    \label{1}
\end{figure*}

\begin{figure*}
    \centering
    \subfloat[]{
        \label{fig:volt1cnv}
        \includegraphics[width=0.482\linewidth, bb= 60 240 740 750, clip]{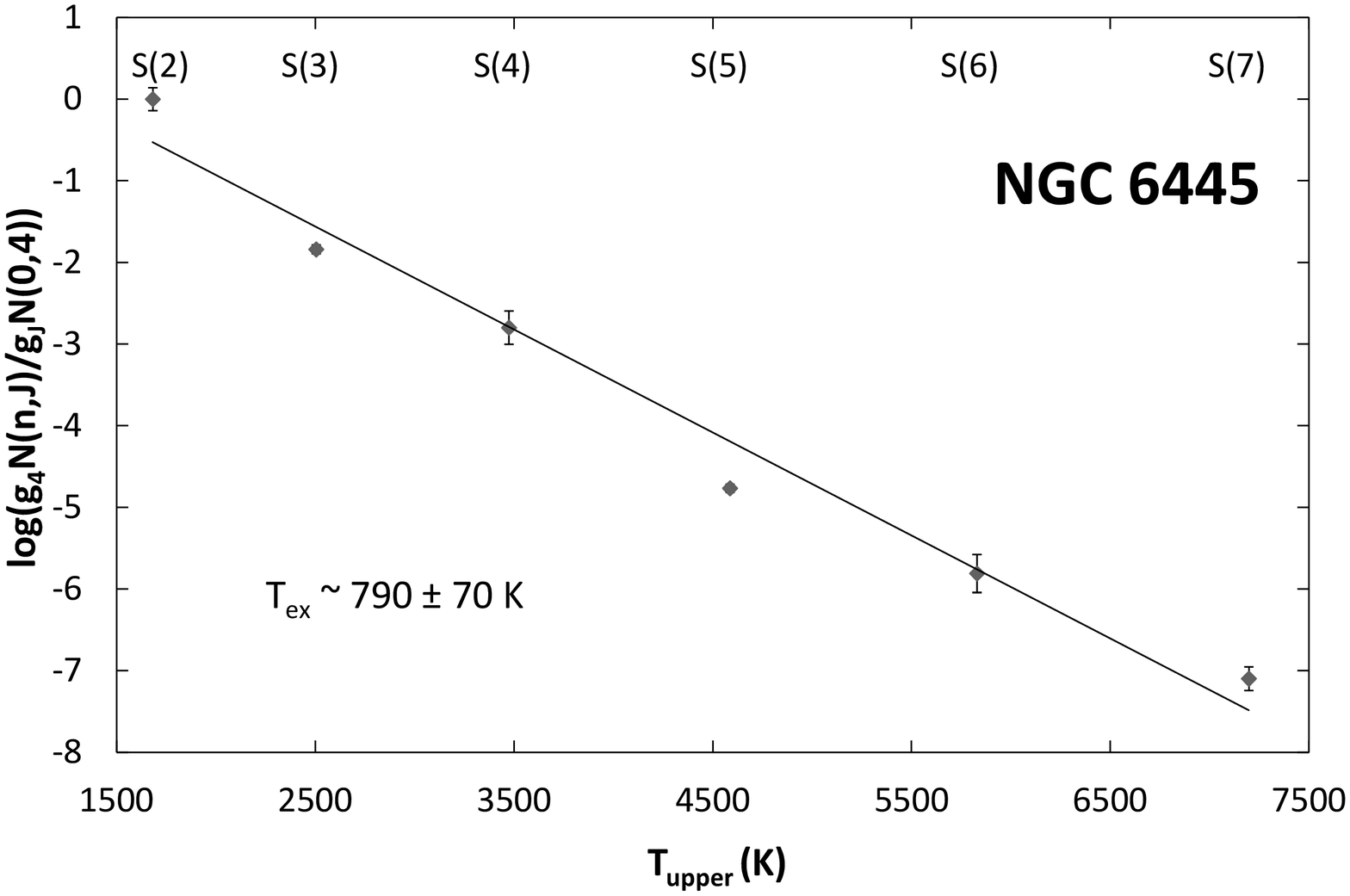}}
    \hspace{0.01\linewidth}
    \subfloat[]{
        \label{fig:crt1cnv}
        \includegraphics[width=0.482\linewidth, bb= 60 240 740 750, clip]{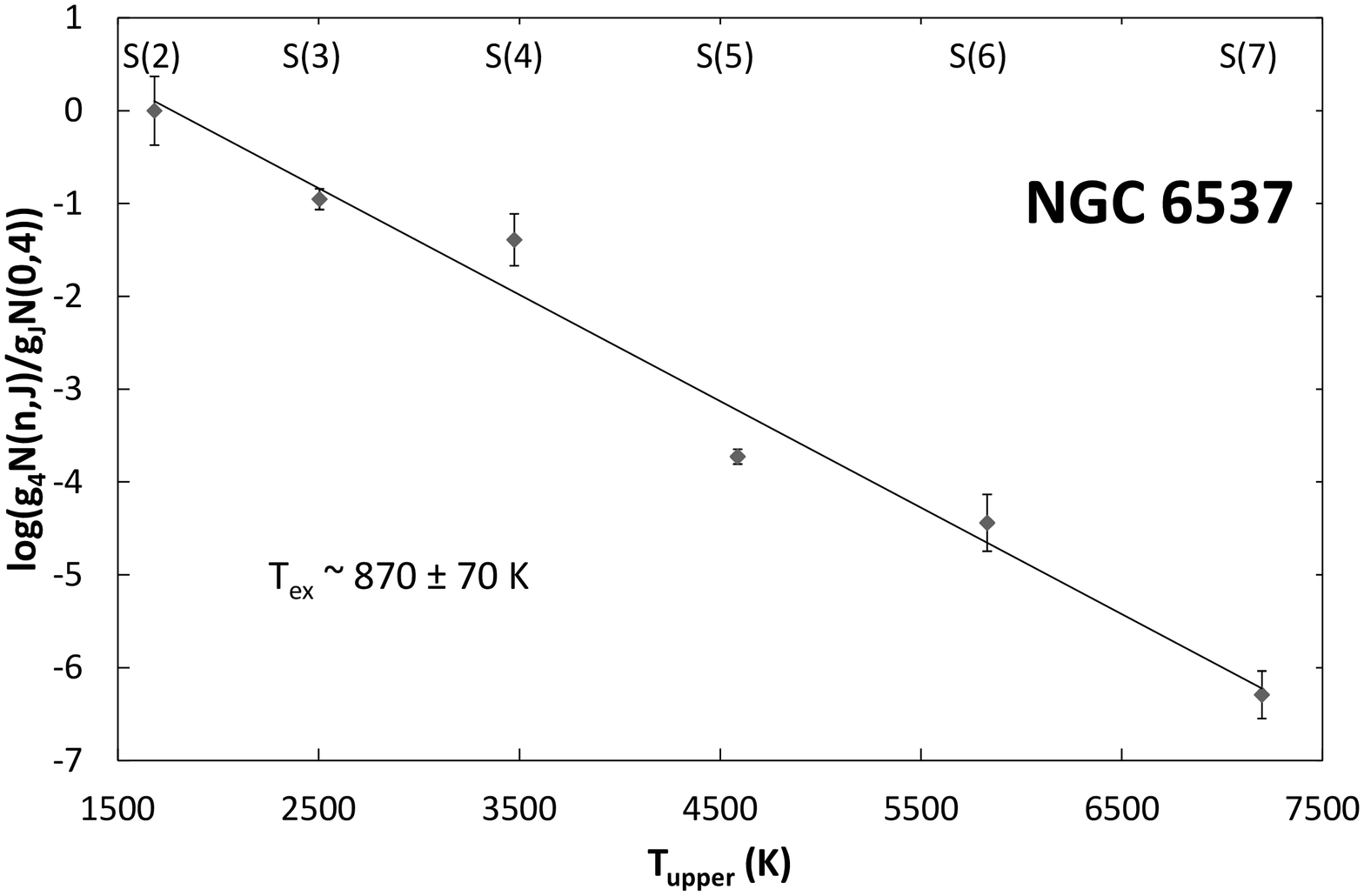}}\\
    \subfloat[]{
        \label{fig:volt1cnv}
        \includegraphics[width=0.482\linewidth]{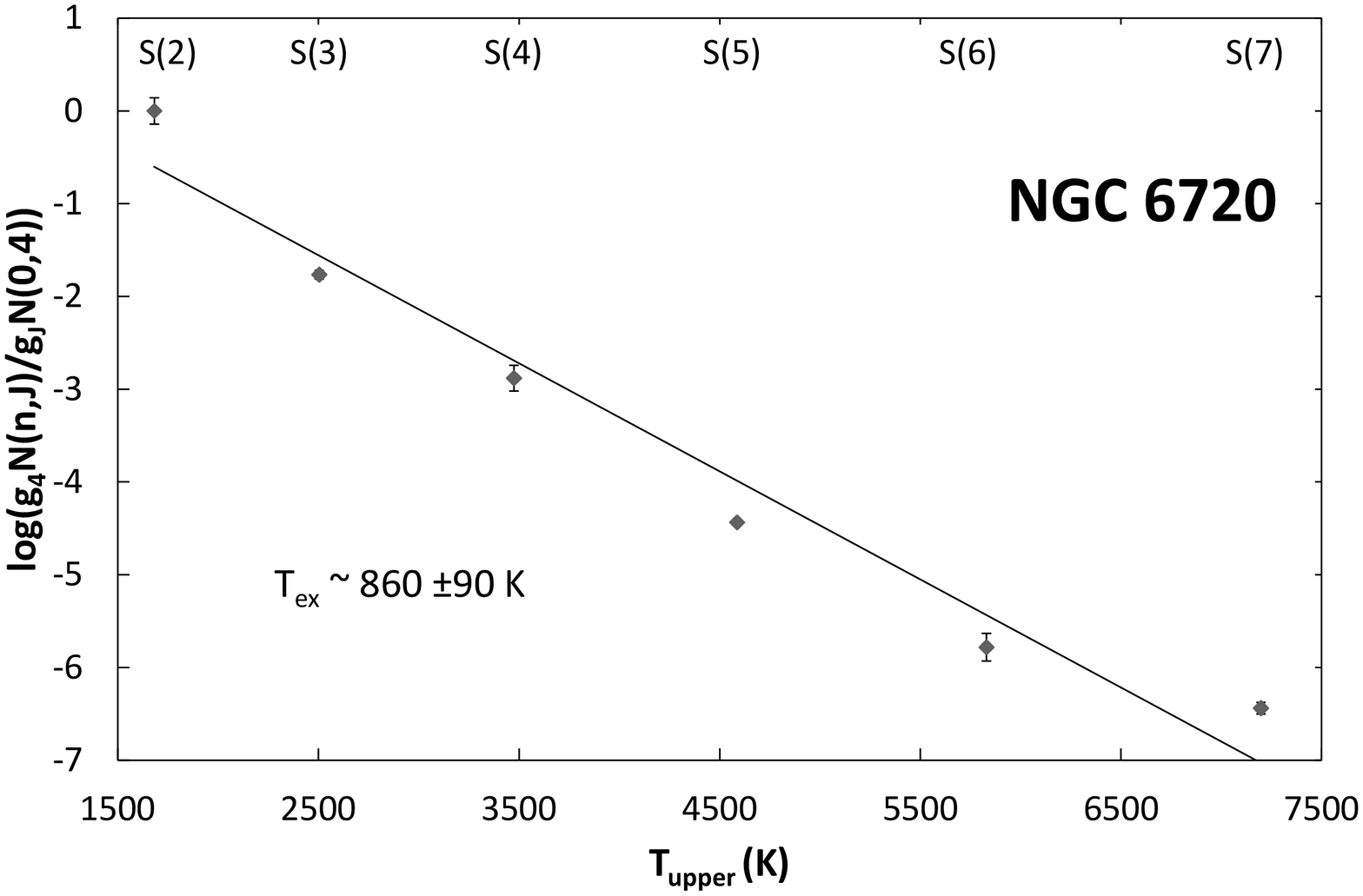}}
    \hspace{0.02\linewidth}
    \subfloat[]{
        \label{fig:crt1cnv}
        \includegraphics[width=0.482\linewidth]{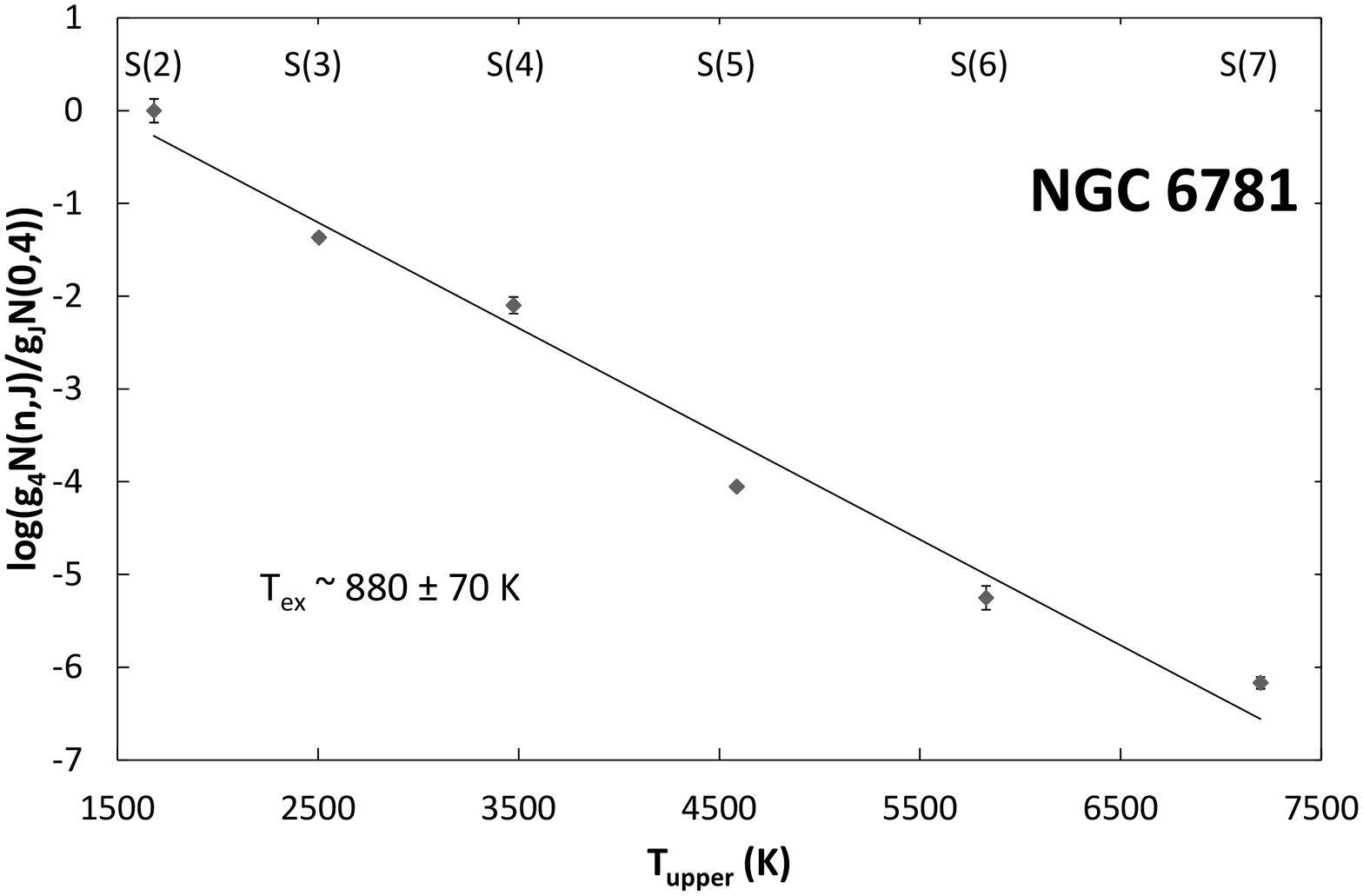}}\\
    \subfloat[]{
        \label{fig:volt1cnv}
        \includegraphics[width=0.482\linewidth]{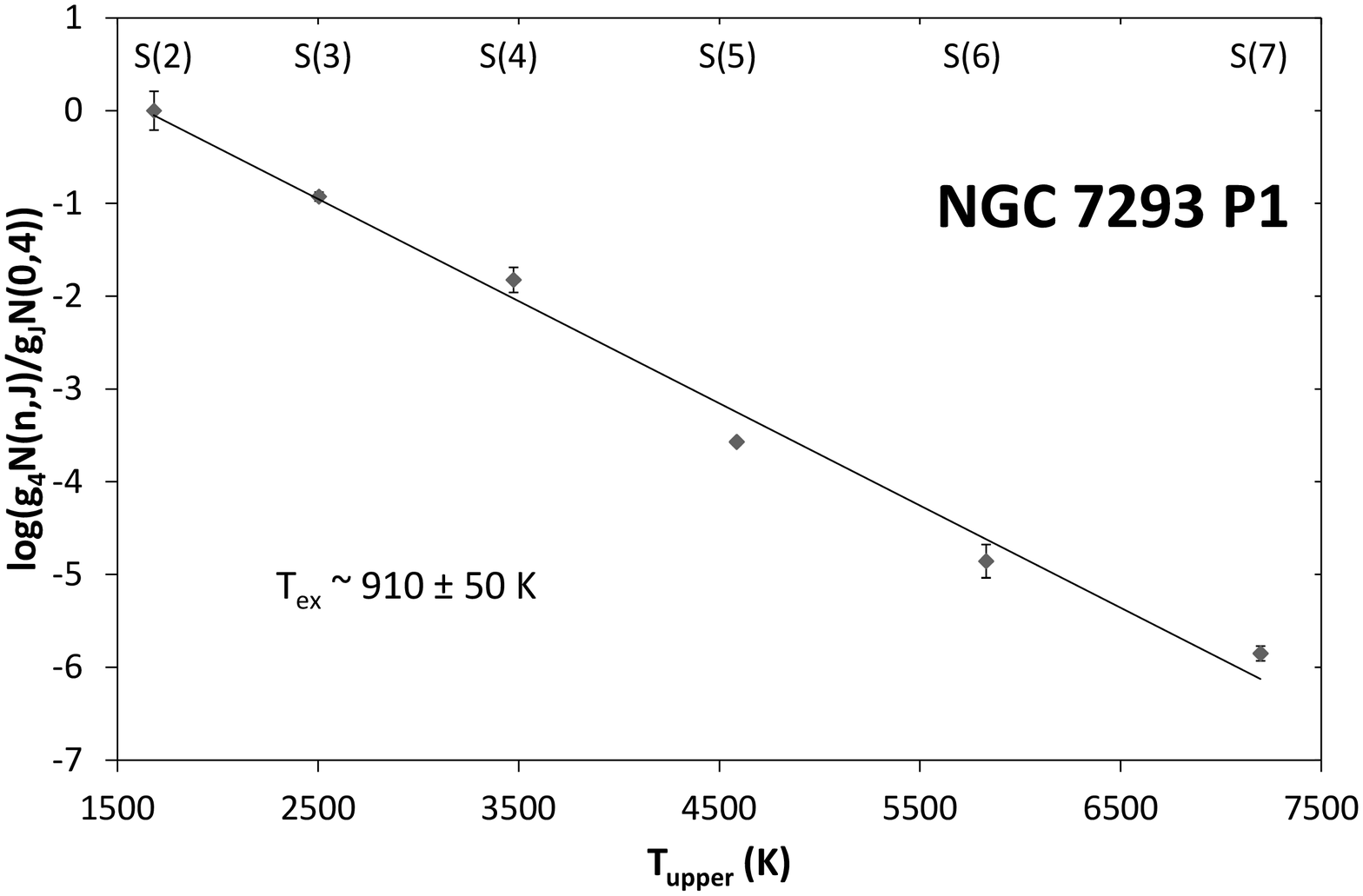}}
    \hspace{0.02\linewidth}
    \subfloat[]{
        \label{fig:crt1cnv}
        \includegraphics[width=0.482\linewidth]{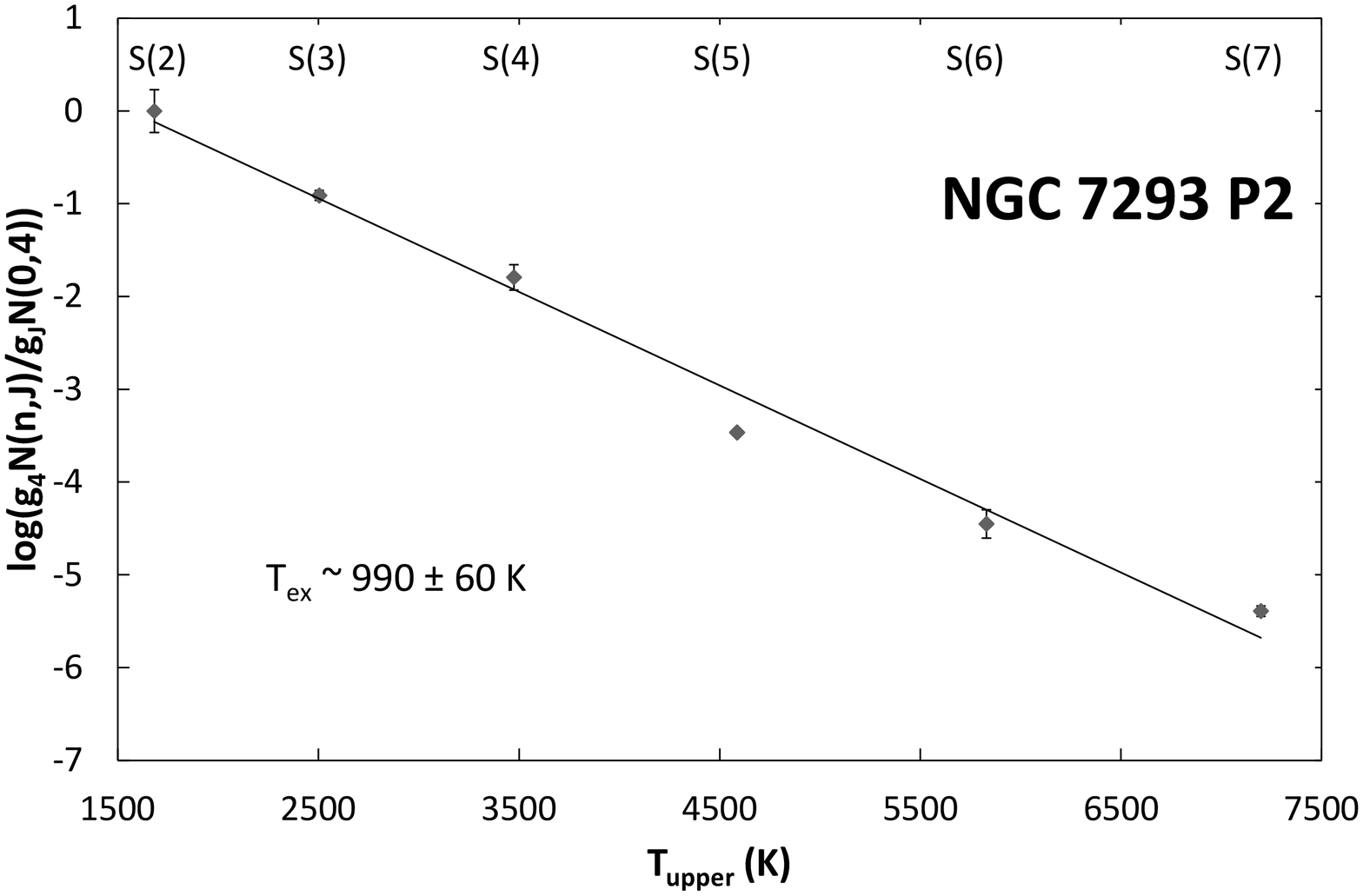}}
    \caption{Same as Fig. 14 but for NGC\,6445, NGC\,6537, NGC\,6720, NGC\,6781, and both positions of NGC\,7293.}
    \label{2}
\end{figure*}

\subsubsection{NGC\,7293}

Infrared studies of the Helix Nebula (NGC\,7293) are numerous, including 
those by \citet{Cox}, \citet{Hora2006} and, more recently, the detailed 
works on H$_{2}$ cometary knots by \citet{Matsuura2007,Matsuura2009} and 
\citet{Aleman2011}.  
The first of these investigations reported an H$_{2}$ excitation temperature 
of 900$\pm$50 K for the west edge, while the second one reported mid-IR fluxes 
for the same AORs as our present work (north and south-west) and 
colour-colour diagrams, but did not calculate the H$_{2}$ excitation 
temperature. 

The close similarity of the spectra at positions 1 and 2 for NGC\,7293 
(Figures \ref{NGC7293P1Spectra} and \ref{NGC7293P2Spectra}) confirms the 
symmetry of the nebula.
The ionic transitions of [S~{\sc iv}] and [Ne~{\sc v}] are notably absent 
in these spectra.  
The spectra are otherwise dominated by the S(5), S(3) and S(7) H$_{2}$ 
rotational lines.

\subsection{Line Intensities}

Table \ref{FluxSummary} shows a summary of the H$_{2}$ and ionic emission 
lines and PAH features identified in the mid-IR spectra of all targets.  
The H$_2$ S(5) 6.91 $\mu$m and [Ar~{\sc ii}] 6.99 $\mu$m emission 
lines are at the limit of the spectral resolution at this wavelength, 
$\simeq$0.05 $\mu$m.  
The relative contributions of both lines has been estimated 
using standard {\sc IRAF} routines for line deblending.  
With the only exception of NGC\,6445, the H$_2$ S(5) line is found to be 
much brighter than the [Ar~{\sc ii}] line and the flux of the latter line 
is thus not reported.

Among all the molecular hydrogen transitions, the S(5) and S(3) lines were the most prominent in all objects.  
Many sources show bright forbidden transitions from ionic species.  [Ar~{\sc iii}] and [Ne~{\sc ii}] were identified 
in all sources, and [S~{\sc iv}] was found in all but two,  M\,2-51 and NGC\,7293.  Other ionic emission lines include those of [Ar~{\sc ii}], [Ar~{\sc v}], and [Ne~{\sc v}]. 

The variations in the [Ne~{\sc v}]/[Ne~{\sc ii}] line ratios among the objects in our sample have a clear correlation with the effective temperature of the central stars (Figure \ref{LineRatios}-{\it left}). Only sources whose CSPN are hotter than 150,000 K have [Ne~{\sc v}] $\lambda$14.32 $\mu$m line intensities notably larger than those of the [Ne~{\sc ii}] $\lambda$12.81 $\mu$m line.  
The larger Ne$^{+3}$ ionization potential (97.1 eV) with 
respect to that of neutral Ne (21.6 eV) provides an easy 
explanation.

Figure~\ref{LineRatios}-{\it right} shows a notable correlation 
between the [Ar~{\sc iii}]/[Ne~{\sc ii}] line ratio and the stellar 
effective temperature.  
With the notable exception of NGC\,7293, the [Ar~{\sc iii}]/[Ne~{\sc ii}] 
line ratio declines steadily from 0.8 for the coldest stars 
($T_{\rm eff}\sim80,000$ K) to 0.2 for NGC\,6537 ($T_{\rm eff}\sim250,000$ K).  
This correlation is intriguing, since the ionization potentials required to 
produce the Ar$^{++}$ and Ne$^+$ species, 27.6 and 21.6 eV, respectively, 
seem to imply that this line ratio should increase with increasing effective 
temperature of the central star.  
Indeed, simple {\sc CLOUDY} simulations for standard parameters for 
PNe ($L_\star$ = 1000 $L_\odot$, $N_{\rm e}$ = 1000 cm$^{-3}$, radius = 
10$^{17}$ cm, typical PN abundances) confirm this expectation: the 
[Ar~{\sc iii}]/[Ne~{\sc ii}] line ratio increases by a factor of 4 
in the range of effective temperature of the central stars in the 
sample.

The CEL emissivity has an exponential dependence on the electron 
temperature $\propto T_{\rm e}^{-1/2} exp(-E_{\rm ex}/kT_{\rm e})$, 
where $E_{\rm ex}$ is the excitation energy of a CEL.  
For an IR fine-structure line of a heavy element (e.g., [Ar~{\sc iii}] 
$\lambda$8.99 $\mu$m or [Ne~{\sc ii}] $\lambda$12.81 $\mu$m lines), the 
excitation energy $E_{\rm ex}$ is very small, and thus its emissivity is 
rather insensitive to the electron temperature.  
However, the emissivity of an IR fine-structure line has a density 
dependence, and the ratio of these two IR lines would be sensitive 
to the electron density.  
{\sc cloudy} models to assess these effects show that the 
[Ar~{\sc iii}]/[Ne~{\sc ii}] line ratio decreases as the 
electron density increases.  
Therefore, the observed trend of the line ratio in our PN sample 
can be explained if the sources with higher effective temperatures 
also have high electron density regions as sampled by the 
\emph{Spitzer} aperture.  
Alternatively, the possibility that the change in the 
[Ar~{\sc iii}]/[Ne~{\sc ii}] line ratio could also be 
a result of the relative abundance of argon and neon 
cannot be totally ruled out.

Finally, it must be noted that there is not a clear correlation 
of the ratios of molecular hydrogen and ionic line intensities 
with the stellar effective temperature.  
There is neither a clear correlation with the nebular density.  
Apparently, the mid-IR H$_2$ and ionic line ratios do not depend on the 
nebular size, central star effective temperature, or nebular density.

\section{Molecular Hydrogen Excitation}
\label{H2Excitation}

Within a given electronic level, the different energy levels of the 
H$_2$ molecules are described by the vibrational quantum number $\nu$ 
and rotational quantum number $J$.
The symmetry of the H$_2$ molecule implies that it has no dipole moment, 
making dipole rotational/vibrational transitions in the same electronic state 
forbidden. 
Electric quadrupole transitions may still occur, with selection rules 
$\Delta J$ = 0 or $\pm$2, with 0-0 transitions forbidden, and no strict 
rules for $\Delta \nu$ \citep{Shull}.  
These transitions are labelled according to the change in $\nu$ and 
the final value of $J$ ($\Delta J$ = $-$2, 0, and $+$2 are labelled 
as the $O$, $Q$ and $S$ branches, respectively).

The H$_2$ 0-0 rotational lines lie between 3 and 28 $\mu$m. Since the H$_{2}$ molecule has a very small moment of inertia, the rotational lines are widely spaced and easily separated by low-resolution spectrometers. In the mid-IR region observed by the low-resolution IRS \textit{Spitzer} spectra, the H$_{2}$ lines correspond to the pure rotational transitions $\nu$ = 0-0 S(2) to S(7) within the vibrational and electronic ground states.

The H$_{2}$ molecule can be excited by different mechanisms, including fluorescence and collisional excitation \citep{Habart}. In the first case, the absorption of a UV photon in the Lyman \& Werner bands results in a rotational-vibrational cascade towards the ground state \citep{Black}. In a PN, these UV photons can originate in the central star, but they can also be the result of emission from strong shocks.  Alternatively, when shocks are moderate, the H$_{2}$ transitions arise through thermal excitation \citep{Sternberg}. The different predicted line strengths between thermal collisional and fluorescence excitation can allow us to distinguish between these two mechanisms by means of IR spectroscopy, but this requires information on the line ratios of H$_2$ lines from differing vibrational states $\nu$, because collisions can thermalize the lowest vibrational levels and mimic thermal excitation. Sometimes this happen if the gas is very dense \citep{Black}.

Using an excitation-energy diagram, the column density in a given level 
with rotational state $J$, relative to the $\nu$ = 0-0 S(2) transition 
having $J$=4, is given by

\begin{equation}
\frac{g_{4} N(\nu, J)}{g_{J} N(0,4)} = \exp\bigg\lbrace - \frac{E(\nu, J)-E(0,4)}{k T_\mathrm{ex}}\bigg\rbrace,
\end{equation}

where the quotient on the left-hand side is equivalent to

\begin{equation}
\label{cociente}
\frac{F(\nu',J')\nu_{0,0 S(2)} A_{0,0 S(2)} g_{4}}{F(0,4) \nu_{\Delta\nu,\Delta J} A_{\nu',J' \rightarrow \nu'',J''} g_{J}}.
\end{equation}

In this case, $F(\nu',J')$ is the observed line flux, and 
$A_{\nu',J' \rightarrow \nu'',J''}$ is the transition probability 
\citep{Hora1994}.
We have used this expression in conjunction with the molecular parameters 
by \citet{1977ApJS...35..281T} and \citet{Dubrowski} and summarized by 
Darren L. DePoy 
(\url{http://www.astronomy.ohio-state.edu/~depoy/research/observing/molhyd.htm}), to determine the population trends illustrated in the excitation diagrams of each PN.

\subsection{H$_{2}$ Excitation Temperatures}

We used the integrated flux of each H$_{2}$ line from the spectrum of 
each object to calculate the population levels according to the procedure 
outlined in the previous section.  
The population diagram showing the log of the quotient in equation 
\ref{cociente} against the upper temperature for the energy level of 
each transition is shown in Figures~\ref{1} and \ref{2}. 
The data points in these figures all show a linear trend.  
Since all the data points correspond to a unique transition $\nu$, 
this linear trend cannot be directly interpreted as a thermal 
distribution of the population of the energy levels of the H$_2$ 
molecules.  
At any rate, the negative inverse of the slope of the linear fits 
can be used to measure the rotational H$_{2}$ excitation temperature, 
$T_\mathrm{ex}$(rot).

Table~\ref{ResultTable} shows a summary of the rotational excitation 
temperatures obtained for each object of our survey.  
An average $T_\mathrm{ex}$(rot) of $\sim$900$\pm$70 K was found for 
the objects in the survey, with no noticeable differences in temperature among the objects in our sample.  

The regions covered by the slits include very different morphological features (innermost regions, envelopes, bipolar lobes) with different expansion velocities and levels of exposure to UV radiation. This excitation temperature is otherwise similar to that found in NGC\,6302, using also H$_2$ mid-IR emission lines \citep{Matsuura2005}.

\section{Discussion}
\label{Discussion}

Our analysis of archival \emph{Spitzer} IRS SL1 and SL2 spectra of a sample of PNe has identified ionic transitions including [Ar~{\sc iii}], [Ne~{\sc ii}], [Ne~{\sc v}], and [S~{\sc iv}].  
The detection of emission lines from species of neon and argon imply that the progenitor stars of the PNe were all at least second-generation stars. The [Ne~{\sc v}] is dominant in sources with hot central stars, which is an effect of the high ionizing power of their stellar spectra. This high ionization line particularly dominates the spectra of NGC\,2440, 
NGC\,6445, and NGC\,6537, with the hottest CSPN in the sample.  
These nebulae also present bipolar morphologies.

The decline of the [Ar~{\sc iii}]/[Ne~{\sc ii}] line ratio with stellar 
effective temperature may be interpreted as a true decline of the Ar/Ne 
abundance ratio.
This can reflect the synthesis of neon during the late evolutionary stages 
of the progenitor stars of those PNe in our sample with high-temperature 
CSPN.  
High-temperature CSPN have been associated to Type~I PNe 
descending from the most massive intermediate-mass stars 
\citep{Corradi95}.  
The study of the production of $\alpha$-elements in a large sample of 
Galactic PNe reveals that neon can be indeed produced in type I PNe 
\citep{Milingo}, but these theoretical studies do not claim a tight 
correlation between the production of $\alpha$-elements and the 
initial mass of the star.  
Alternatively, the line ratio decline can be coupled with increasing 
nebular densities, resulting also from the large mass ejection and 
fast stellar evolution of the most massive PN progenitors.

The broad PAH feature centred at 11.30 $\mu$m is detected in a number of PNe in this sample. The feature is  present in the spectra of NGC\,2440, NGC\,3132, NGC\,6445, NGC\,6537 and NGC\,6781. 
There is a possible detection in NGC\,2346, but it is absent in M\,2-51, NGC\,2818, NGC\,6720, NGC\,6072 
and NGC\,7293; this latter case concurs with \citet{Hora2006}. 
The presence of this feature is typically interpreted as indicative 
of C-H bending \citep{Fleming} after the absorption of UV photons 
\citep{LegerPuget,Allamandola,Kwok}, as it is found to be weaker in 
non-C rich PNe \citep{VolkCohen}.

Finally, our analysis of NGC\,7293 is consistent with that reported by \citet{Hora2006} regarding the absence of PAH emission.  These authors argue that the PAH emission could have been missed due to the small size of the slits compared to the extension of the object.  Meanwhile, \citet{Cox} reported that the absence of PAHs is consistent with the evolved nature of NGC\,7293, having dissociated or dispersed all of its dust grains.The H$_2$ is shock-excited, with shock velocities of a few tens km~s$^{-1}$ \citep{Hora2006}.  

\begin{table}
\centering
\caption{H$_{2}$ rotational excitation temperatures.}\label{ResultTable}
\begin{tabular}{@{}lccll@{}}
\toprule
\multicolumn{1}{l}{Object} & \multicolumn{2}{c}{T$_\mathrm{ex}$ H$_{2}$} & Ref. & Notes \\
\multicolumn{1}{c}{}       & This work   & Literature   &   & \\
                           &   (K)       &    (K)       &   &  \\ \midrule
M\,2-51                    & 900$\pm$90  & $\dots$      &   & \\
NGC\,2346                  & 950$\pm$110 & 1260$\pm$200 &   & \\
NGC\,2440                  & 830$\pm$80  & 1070$\pm$200 & 1 & North lobe \\
                           &             & 2020$\pm$200 &   & East lobe  \\
NGC\,2818                  & 850$\pm$50  & $\dots$      &   & \\
NGC\,3132                  & 900$\pm$50  & $\dots$      &   & \\
NGC\,6072                  & 890$\pm$60  & $\dots$      &   & \\
NGC\,6445                  & 790$\pm$70  & 1520         & 2 & \\
NGC\,6537                  & 870$\pm$70  & 3600         & 2 & Inner region \\
NGC\,6720                  & 860$\pm$90  & 1250$\pm$100 & 1 & Ring \\
                           &             & 2000$\pm$200 &   & Halo \\
NGC\,6781                  & 880$\pm$70  & $\dots$      &   & \\ 
NGC\,7293 P1 	             & 900$\pm$50  & 900          & 3 & \\ 
NGC\,7293 P2		           & 990$\pm$60  &              &   & \\ \bottomrule
\end{tabular}              
\footnotesize{
References:
(1) \citealp{Hora1999}; (2) \citealp{Davis}; 
(3) \citealp{Cox}.}
\end{table}

\subsection{Previous Determination of H$_2$ Excitation Temperatures}

The H$_2$ rotational excitation temperatures derived in this work are all in the range of 900 K.  Among the nebulae in our sample, NGC\,2346, NGC\,2440, and NGC\,6720 were included in the near-IR spectroscopic study of upper vibrational levels of the H$_2$ molecule carried out by \citet{Hora1999}.  In that paper, NGC\,2346 is considered to be UV-excited and an H$_{2}$ rotational excitation temperature of 1260$\pm$200 K is reported.  This value is larger that the one reported here, 950$\pm$110 K, but within the uncertainties.  The temperatures derived for different regions of NGC\,2440 and NGC\,6720 range from 1070$\pm$200 K to 2020$\pm$200 K for the former and from 1240$\pm$100 K to 2000$\pm$200 K for the latter.  \citet{Hora1999} suggests that all these regions (see Table~4) are UV excited, except probably the halo of NGC\,6720.  
The temperature in the north lobe of NGC\,2440 and ring of NGC\,6720 would be consistent with our results, but not those in the hottest regions.

Other PNe with reported H$_2$ excitation temperatures are NGC\,6445 and NGC\,6537 \citep[1520 K and 3600 K, respectively,][]{Davis} and NGC\,7293 \citep[900 K,][]{Cox}. 
The most discrepant temperature is that of NGC\,6537, but we note that \citet{Davis} derived a value for the innermost nebular regions of this extremely high excitation PN, whereas the mid-IR spectrum presented here traces the outer bipolar lobes.

NGC\,6781 has been reported to exhibit strong [Ar~{\sc iii}], [S~{\sc iv}], and [Ne~{\sc ii}] emission in the internal nebular shell that drops off suddenly outside its borders \citep{Phillips2011}.  Our study confirmed the line intensity ratios of these ionic lines, whereas the H$_2$ excitation temperature provided by \citet{Phillips2011} are in good agreement with the value of 880 K derived in this work.

\section{Summary \& Conclusions}

We present an investigation of archival \emph{Spitzer} IRS spectra of a 
sample of PNe to search for mid-IR H$_2$ lines.  
Among the original sample of 14 PNe with useful \emph{Spitzer} IRS SL 
observations known to exhibit H$_2$ emission lines in the near-IR, 11 PNe 
show detectable mid-IR H$_2$ lines.  
This result increases the number of known PNe with mid-IR H$_2$ emission 
lines from three (NGC\,6302, NGC\,6781, and NGC\,7293) to twelve.

The spectral analysis detects all the H$_2$ 0-0 transitions from the 
S(2) to the S(7) lines.  
Among these lines, the H$_2$ 0-0 S(3) $\lambda$9.66 $\mu$m line is the 
brightest.  
The analysis of the population distribution in the H$_2$ molecules of 
these PNe reveals a T$_\mathrm{ex}$(rot) excitation temperature of 900 
K for all of them.  
The conspicuous uniformity of the excitation temperature is intriguing, 
because the observations trace different morphological features of a 
non-uniform sample of PNe with different morphologies and evolutionary 
stages.  
Near-IR studies are needed to confirm these excitation temperatures and to 
diagnose the excitation mechanism of H$_2$ in these PNe.

Ionic lines of [Ar~{\sc iii}], [S~{\sc iv}], [Ne~{\sc ii}], and [Ne~{\sc v}] 
are also detected.  
Their line intensities show the expected positive correlation of the 
[Ne~{\sc v}]/[Ne~{\sc ii}] line ratio with the CSPN effective temperature.  
On the other hand, the anti-correlation between the 
[Ar~{\sc iii}]/[Ne~{\sc ii}] and CSPN effective temperature 
is not expected and may imply
changes in the elemental ratios that suggest that PNe with 
hottest CSPN, probably descending from more massive progenitors, 
are able to produce neon.  
Alternatively, the more massive progenitors produce regions of 
higher density that reduce the [Ne~{\sc v}]/[Ne~{\sc ii}] line 
ratio.

\section*{Acknowledgements}
\label{sec:acknowledgements}

We would like to thank an anonymous referee for several perceptive remarks.
The authors thank V. Guzm\'an-Jim\'enez for her advice and insight into 
the CUBISM package.
HM is grateful for CONACyT scholarship 478329/276214.
GRL acknowledges support from CONACyT (grant 177864), CGCI, PROMEP and SEP 
(Mexico).
JAT acknowledges support by the CSIC JAE-Pre student grant 2011-00189.
J.A.T.\ and M.A.G.\ are supported by the Spanish MICINN (Ministerio de 
Ciencia e Innovaci\'on) grant AYA 2011-29754-C03-02 and AYA 2014-57280-P 
co-funded with FEDER funds. 
This research has made use of the NASA/IPAC Infrared Science Archive, 
which is operated by the Jet Propulsion Laboratory at the California Institute 
of Technology, under contract with the National Aeronautics and Space 
Administration.
The Digitized Sky Survey images were produced at the Space Telescope 
Science Institute under U.S. Government grant NAG W-2166. 
The images of these surveys are based on photographic data obtained 
using the Oschin Schmidt Telescope on Palomar Mountain and the UK Schmidt 
Telescope. 
The plates were processed into the present compressed digital form with 
the permission of these institutions.

\end{document}